\documentclass[prd,preprint,superscriptaddress,preprintnumbers,eqsecnum,showpacs,nofootinbib,nobibnotes,noeprint]{revtex4-1}
\usepackage[linkcolor=blue,citecolor=blue,urlcolor=blue,colorlinks=true,breaklinks]{hyperref} 

\usepackage{amsfonts,bm}
\usepackage{amsfonts,amssymb,amsmath}
\usepackage{graphicx}
\usepackage{paralist}
\usepackage{enumitem}
\usepackage[english]{babel}
\usepackage{slashed}
\usepackage[usenames]{xcolor}
\usepackage{mathtools}
\usepackage[makeroom]{cancel}
\usepackage{color}

\def\srm#1{{\rm{\scriptscriptstyle #1}}}

\newcommand{\be}{\begin{equation}}
\newcommand{\bea}{\begin{eqnarray}}
\newcommand{\ee}{\end{equation}}
\newcommand{\eea}{\end{eqnarray}}

\def\1eq#1{Eq.~(\ref{#1})}

\def\2eqs#1#2{Eqs.~(\ref{#1}) and~(\ref{#2})}
\def\3eqs#1#2#3{Eqs.~(\ref{#1}),~(\ref{#2}) and~(\ref{#3})}

\def\fig#1{Fig.~\ref{#1}}

\def\ie{{\it i.e.}, }
\def\eg{{\it e.g.}, }

\newcommand{\p}{\mathbf{p}}
   
\def\g{\Gamma}

\def\s#1{{\scriptscriptstyle #1}}

\newcommand{\fatg}{{\rm{I}}\!\Gamma}

\newcommand{\Cfat}{{\mathbb C}}

\newcommand{\Rc}[1]{ V_{#1} }

\newcommand{\Vc}[1]{ {\mathbb V}_{#1} }

\newcommand{\bV}{{\overline V}\vphantom{V}}
\newcommand{\qV}[1]{ \bV_{#1} }

\newcommand{\qG}[1]{ S_{#1} }

\begin{document}

\title{Schwinger poles of the three-gluon vertex: 
\\ symmetry and dynamics}

\author{A.~C. Aguilar}
\affiliation{\mbox{University of Campinas - UNICAMP, Institute of Physics ``Gleb Wataghin'',} \\
13083-859 Campinas, S\~{a}o Paulo, Brazil}

\author{M.~N. Ferreira}
\affiliation{\mbox{Department of Theoretical Physics and IFIC, 
University of Valencia and CSIC},
E-46100, Valencia, Spain}

\author{B.~M. Oliveira}
\affiliation{\mbox{University of Campinas - UNICAMP, Institute of Physics ``Gleb Wataghin'',} \\
13083-859 Campinas, S\~{a}o Paulo, Brazil}

\author{J. Papavassiliou}
\affiliation{\mbox{Department of Theoretical Physics and IFIC, 
University of Valencia and CSIC},
E-46100, Valencia, Spain}

\author{L.~R. Santos}
\affiliation{\mbox{University of Campinas - UNICAMP, Institute of Physics ``Gleb Wataghin'',} \\
13083-859 Campinas, S\~{a}o Paulo, Brazil}

\begin{abstract}

The implementation of the Schwinger mechanism 
endows gluons with a 
nonperturbative mass through the formation of special massless 
poles in the fundamental QCD vertices;
due to their longitudinal character, these poles do not cause divergences in on-shell amplitudes, 
but induce detectable effects in 
the Green's functions of the theory. 
Particularly important in this 
theoretical setup is the three-gluon vertex,
whose pole content extends beyond the minimal structure 
required for the generation of a gluon mass. 
In the present work we analyze these additional pole patterns by means of two distinct, but ultimately equivalent, 
methods: the Slavnov-Taylor identity satisfied by the 
three-gluon vertex, and the 
nonlinear Schwinger-Dyson equation that 
governs the dynamical evolution of this vertex. 
Our analysis reveals that the 
Slavnov-Taylor identity imposes strict model-independent 
constraints on the associated residues, 
preventing them 
from vanishing.
Approximate versions of these constraints are subsequently recovered from the Schwinger-Dyson equation, 
once the elements responsible for the 
activation of the Schwinger mechanism have been duly incorporated. The excellent coincidence between the two 
approaches exposes a profound connection between symmetry and dynamics, and 
serves as a nontrivial 
self-consistency test of this particular 
mass generating scenario. 

\end{abstract}


\maketitle

\section{\label{sec:intro} Introduction}

The emergence of a gluon mass~\cite{Cornwall:1981zr,Bernard:1981pg,Bernard:1982my,Parisi:1980jy,Donoghue:1983fy,Mandula:1987rh,Cornwall:1989gv,Wilson:1994fk,Aguilar:2008xm} through the 
action of the Schwinger mechanism~\mbox{\cite{Schwinger:1962tn,Schwinger:1962tp}} 
represents a prime example of how mass may emanate from interaction~\cite{Roberts:2020hiw}. Indeed, 
the most appealing attribute of this mechanism 
is that it arises entirely from the 
underlying dynamics, 
without the slightest modification of the 
fundamental Lagrangian that defines the theory,
and, most importantly, leaving the local gauge symmetry intact~\cite{Papavassiliou:2022wrb,Ferreira:2023fva}.

The cornerstone of the Schwinger mechanism is the nonperturbative formation of colored 
composite excitations with vanishing mass in the vertices of the theory~\cite{Jackiw:1973tr,Jackiw:1973ha,Eichten:1974et,Poggio:1974qs,Smit:1974je,Cornwall:1973ts,Cornwall:1979hz}, and especially in the three-gluon vertex, $\fatg_{\alpha\mu\nu}(q,r,p)$~\cite{Cornwall:1981zr,Aguilar:2008xm,Aguilar:2011xe,Binosi:2012sj,Ibanez:2012zk}; for a variety of different approaches, see~\cite{Kondo:2001nq,Braun:2007bx,Fischer:2008uz,Campagnari:2010wc,Tissier:2010ts,Serreau:2012cg,Pelaez:2014mxa,Siringo:2015wtx,Eichmann:2021zuv,Horak:2022aqx}.

A special subset of these massless poles is
transmitted to the gluon propagator,
$\Delta(q)$, through the coupled 
dynamical equations 
of motion, \ie 
Schwinger-Dyson equations (SDEs)~\cite{Roberts:1994dr,Alkofer:2000wg,Fischer:2006ub,Roberts:2007ji,Binosi:2009qm,Cloet:2013jya,Aguilar:2015bud,Huber:2018ned,Papavassiliou:2022wrb,Ferreira:2023fva}, 
triggering finally 
its saturation at the origin, 
\mbox{$\Delta^{-1}(0) =m^2 >0$} \mbox{\cite{Aguilar:2011xe,Binosi:2012sj,Ibanez:2012zk,Aguilar:2017dco,Binosi:2017rwj}}. 

Due to the special dynamical details 
governing their formation, 
the massless poles of the three-gluon vertex are   
{\it longitudinally coupled}~\mbox{\cite{Jackiw:1973tr,Jackiw:1973ha,Eichten:1974et,Poggio:1974qs,Smit:1974je,Cornwall:1973ts,Cornwall:1979hz}}, 
\ie they correspond to tensorial 
structures of the general form 
$q_\alpha/q^2$, $r_{\mu}/r^2$, 
and $p_{\nu}/p^2$.
As a result, they are not directly 
detectable in on-shell amplitudes, nor
in lattice simulations of the  
corresponding correlation functions~\cite{Parrinello:1994wd,Alles:1996ka,Parrinello:1997wm,Boucaud:1998bq,Cucchieri:2006tf,Maas:2007uv,Cucchieri:2008qm,Athenodorou:2016oyh,Duarte:2016ieu,Boucaud:2017obn,Sternbeck:2017ntv, Vujinovic:2018nqc,Boucaud:2018xup,Aguilar:2019uob,Aguilar:2021lke,Pinto-Gomez:2022brg,Catumba:2021hng,Catumba:2021yly};
nonetheless, their effects are. 
Thus, in addition to causing the  
infrared saturation of the gluon propagator, 
the form factor 
$\Cfat(q)$
associated with the pole 
induces a smoking-gun modification (``displacement'')
to  
the Ward identity of the three-gluon vertex~\cite{Aguilar:2021uwa,Aguilar:2022thg,NarcisoFerreira:2023kak,Ferreira:2023fva}. Most importantly, 
the nonvanishing of   $\Cfat(q)$ has 
been unequivocally 
confirmed in~\cite{Aguilar:2022thg}, 
through the suitable 
combination of key inputs obtained  
from lattice QCD~\cite{Bogolubsky:2009dc,Boucaud:2018xup,Aguilar:2021lke,Aguilar:2021okw}.

This encouraging result motivates the further detailed scrutiny 
of the key features that the Schwinger mechanism induces 
in the three-gluon vertex. 
The main purpose of the present work 
is to carry out an extensive study of the full pole content 
of this vertex, determine the
structure and role of the main 
components, and 
expose the delicate interplay 
between symmetry and dynamics 
that prompts their appearance. In that sense, 
our analysis provides a nontrivial 
confirmation of the internal consistency of this rather elaborate mass generating approach. 

The dynamics of the pole formation 
are encoded in the 
nonlinear SDE that 
controls the evolution of 
$\fatg_{\alpha\mu\nu}(q,r,p)$.
In their primordial manifestation, 
the massless poles arise 
as bound states of a particular kernel 
appearing in the skeleton expansion 
of this SDE~\cite{Aguilar:2011xe,Aguilar:2011xe,Binosi:2012sj,Ibanez:2012zk,Binosi:2017rwj,Aguilar:2017dco}; they  
are simple, 
of the type $1/q^2$, $1/r^2$, and $1/p^2$. When inserted into the  
SDE for $\Delta(q)$, only the 
pole in the direction of $q$ 
is relevant for the 
generation of the gluon mass,
which is  expressed as 
an integral over 
the residue of this pole.
However, due to the nonlinear nature of 
the vertex SDE,
these ``primary'' poles give rise to additional 
``secondary'' structures, 
corresponding to 
mixed double poles, of the general type 
$1/q^{2}r^{2}, 1/q^{2}p^{2}, 
1/r^{2}p^{2}$. In the Landau gauge, 
these poles are inert as far as mass generation is concerned; 
however, their presence is instrumental for the self-consistency 
of the entire approach, and in particular for 
preserving the fundamental relations that arise from 
the Becchi-Rouet-Stora-Tyutin (BRST) symmetry~\cite{Becchi:1975nq,Tyutin:1975qk} of the gauge-fixed 
Yang-Mills Lagrangian.

Indeed, 
the emergence of mixed poles 
finds its most compelling justification when the Slavnov-Taylor identity (STI)~\cite{Taylor:1971ff,Slavnov:1972fg} 
of the three-gluon vertex~\cite{Marciano:1977su,Ball:1980ax,Davydychev:1996pb,Gracey:2019mix} 
is invoked.
In its abelianized version, with the contributions of the ghost sector  
switched off, 
this STI states that 
\mbox{$q^\alpha \fatg_{\alpha\mu\nu}(q,r,p) = 
P_{\mu\nu}(p)\Delta^{-1}(p) - 
P_{\mu\nu}(r)\Delta^{-1}(r)$},
where $P_{\mu\nu}(q) = 
g_{\mu\nu} - q_{\mu}q_{\nu}/q^2$
is the standard projection operator. 
Let us now assume that 
the gluon propagator 
is infrared finite, \ie $\Delta^{-1}(0) = m^2$.
Then, in the limit $p^2\to 0$ or $r^2\to 0$, 
the r.h.s.\,\,of the STI displays longitudinally coupled massless poles,  
$p_{\mu}p_{\nu}/p^2$ and $r_{\mu}r_{\nu}/r^2$, whose residue is $m^2$. Consequently, 
self-consistency requires that, 
in the same kinematic limits, the l.h.s. should 
exhibit the exact same pole structure, \ie  $\fatg_{\alpha\mu\nu}(q,r,p)$ 
must contain mixed poles, of the 
type $q_\alpha p_{\mu}p_{\nu}/q^2p^2$ and 
$q_\alpha r_{\mu}r_{\nu}/q^2r^2$, precisely 
as predicted by the 
vertex SDE.

The exact matching of pole contributions on both sides of the STI
(with the ghost contributions duly restored)
gives rise to a nontrivial relation, 
which expresses the form factors associated with the mixed poles in terms of components that appear on the r.h.s. of the STI. Quite interestingly, 
an approximate form of this special relation may be 
recovered from a truncated version of the vertex SDE. Moreover, 
an  analogous construction reveals that 
the presence of a genuine triple mixed pole, 
of the type $1/q^2p^2r^2$ is \emph{excluded} 
by both the STI and the SDE, being  
effectively reduced to a divergence weaker than a double mixed pole. 
These two exercises are especially illuminating, 
exposing a powerful synergy between symmetry and dynamics: 
whereas the STI (BRST symmetry) imposes relations 
that are valid regardless of the 
dynamical details, 
the SDE (nonlinear dynamics) reproduces them  
thanks to the distinct pole 
content induced by the Schwinger mechanism.

The article is organized as follows. In 
Sec.~\ref{sec:jss} we summarize the most salient 
features of the Schwinger mechanism in QCD, 
commenting on some of its most recent advances. 
Then, in 
Sec.~\ref{sec:background} 
we discuss in detail the pole structure induced to the three-gluon 
vertex when the Schwinger mechanism is activated,
and in particular the appearance of mixed double and triple poles. In Sec.~\ref{sec:basis} we construct a tensor basis for the pole part of the vertex, which makes its Bose symmetry and longitudinal nature manifest, and will be used throughout this work. 
In Sec.~\ref{sec:3g_const} we consider the STI satisfied by the 
three-gluon vertex,
and derive a crucial relation 
for a special kinematic limit of 
the form factor associated with the 
mixed double poles, 
denominated ``residue function''.
Then, 
in Sec.~\ref{V9SDE}, we 
turn to the SDE of the three-gluon vertex, 
and derive, under certain simplifying assumptions,  
an approximate version of the 
aforementioned 
relation 
for the residue function.  
In continuation, in Sec.~\ref{num} we 
compute the residue function using 
as inputs all the components 
entering in that relation. 
Then, in Sec.~\ref{triple}, we demonstrate that 
both the STI and the detailed dynamics 
reduce substantially the strength of  
the triple mixed pole. 
Finally, in Sec.~\ref{conc} we present our discussion and conclusions.

\section{Schwinger mechanism in QCD: General concepts}\label{sec:jss}

In this section we present a brief overview of the 
implementation of the Schwinger mechanism in the context 
of a Yang-Mills theory; for further details, the reader 
is referred to two recent review articles~\cite{Papavassiliou:2022wrb,Ferreira:2023fva}. 

\begin{figure}[t]
\includegraphics[width=1.0\linewidth]{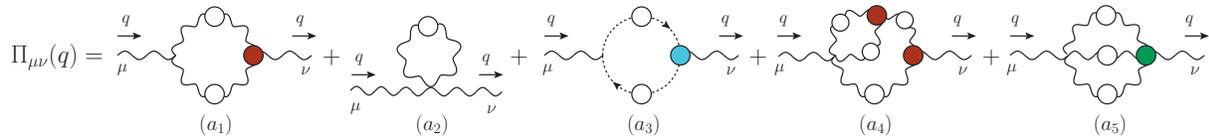}
\caption{The diagrammatic representation  of the gluon self-energy.
The fully-dressed three-gluon, ghost-gluon, 
and four-gluon vertices are depicted 
as red, blue,  and green circles, respectively.
The special analytic structure 
of these vertices induces the poles required for the activation of the Schwinger mechanism.}
\label{fig:gluonself}
\end{figure}

The natural starting point of the discussion is the gluon propagator, \mbox{$\Delta^{ab}_{\mu\nu}(q)=-i\delta^{ab}\Delta_{\mu\nu}(q)$}. In the \emph{Landau gauge} that we employ 
throughout, $\Delta_{\mu\nu}(q)$ assumes 
the completely transverse form 
\be
\Delta_{\mu\nu}(q) = \Delta(q) {P}_{\mu\nu}(q)\,, \qquad {P}_{\mu\nu}(q) := g_{\mu\nu} - q_\mu q_\nu/{q^2}\,.
\label{defgl}
\ee
In the continuum, 
the momentum evolution of the function $\Delta(q)$  
is determined by the corresponding SDE 
(Minkowski space), 
\be
\Delta^{-1}(q) = q^2  + i \Pi(q) \,,
\label{glSDE}
\ee
where $\Pi(q)$ is the scalar form factor 
of the gluon self-energy,
\be \Pi_{\mu\nu}(q) =\Pi(q){P}_{\mu\nu}(q) \,,
\ee
depicted diagrammatically in \fig{fig:gluonself}. 
Note that 
the fully-dressed vertices, $\fatg$,  
of the theory 
enter in the diagrams defining 
$\Pi_{\mu\nu}(q)$. 
In addition, it is convenient to introduce 
the dimensionless 
vacuum polarization, ${\bf \Pi}(q)$, defined as 
$\Pi(q) = q^2 {\bf \Pi}(q)$, such that  \mbox{$\Delta^{-1}({q})=q^2 [1 + {\bf \Pi}(q)]$}.

The basic premise underpinning the Schwinger mechanism may be expressed 
as follows:
if ${\bf \Pi}(q)$ develops a pole with positive residue at $q^2=0$ 
(\emph{massless pole}),
the gauge boson (gluon) acquires a mass,  
even if the symmetries of the theory do not admit a mass term at the 
level of the fundamental Lagrangian~\mbox{\cite{Schwinger:1962tn,Schwinger:1962tp}}. 
In particular, the appearance of such a pole triggers  
the basic sequence (Euclidean space) 
\be
\lim_{q \to 0} {\bf \Pi}(q) = m^2/q^2 \,\,\Longrightarrow \,\,\lim_{q \to 0} \,\Delta^{-1}(q) = \lim_{q \to 0} \,(q^2 + m^2) \,\,\Longrightarrow \,\,\,
\Delta^{-1}(0) = m^2 \,,
\label{schmech}
\ee
where the residue of the pole acts as the effective 
squared gluon mass, $m^2$. 

The pivotal result captured by \1eq{schmech} invites the 
natural question of what makes ${\bf \Pi}(q)$ 
exhibit massless poles in the first place. 
In the case of four-dimensional Yang-Mills theories, such as QCD,  
the answer to this question is that these poles 
are transmitted to  ${\bf \Pi}(q)$ by the 
fully-dressed vertices that appear in the 
diagrammatic expansion of the gluon self-energy~\cite{Cornwall:1981zr,Aguilar:2006gr,Aguilar:2008xm,Papavassiliou:2022wrb,Ferreira:2023fva}, see \fig{fig:gluonself}.
The poles of the vertices are produced dynamically,
when elementary fields 
(\eg two gluons, two ghosts, or three gluons) 
merge to create composite colored scalars with 
vanishing masses~\mbox{\cite{Jackiw:1973tr,Jackiw:1973ha,Eichten:1974et,Poggio:1974qs,Smit:1974je,Cornwall:1973ts,Cornwall:1979hz}}.  
These processes are controlled by 
appropriate bound-state equations, 
analogous to the standard 
Bethe-Salpeter equations (BSEs)~\cite{Salpeter:1951sz,Nakanishi:1969ph}; they  
arise as special kinematic limits 
($q\to 0$) of the 
SDEs governing the various vertices~\mbox{\cite{Aguilar:2011xe,Ibanez:2012zk,Aguilar:2017dco,Binosi:2017rwj}}. 
The residues of the vertices 
are 
functions of 
the remaining kinematic variables; 
when 
convoluted 
with the rest of the components 
comprising the gluon SDE, they account for the 
final residue, 
$m^2$,
that one identifies as the squared 
gluon mass
in \1eq{schmech}~\mbox{\cite{Aguilar:2011xe,Ibanez:2012zk,Aguilar:2017dco,Binosi:2017rwj}}.

To elucidate how 
a contribution to the total gluon mass emerges from 
diagram $(a_1)$ in \fig{fig:gluonself}, 
consider the 
three-gluon vertex 
\mbox{$\fatg^{abc}_{\alpha\mu\nu}(q,r,p) = gf^{abc}\fatg_{\alpha\mu\nu}(q,r,p)$}, 
where $g$ is the gauge coupling,  
$f^{abc}$ the structure constants of
the SU(3) gauge group, 
and $q+r+p=0$. 
The formation of the poles in the three-gluon vertex may be described by separating  $\fatg_{\alpha\mu\nu}(q,r,p)$ in
two distinct pieces,
\be
\fatg_{\alpha\mu\nu}(q,r,p) = \g_{\alpha\mu\nu}(q,r,p) + V_{\alpha\mu\nu}(q,r,p)\,,
\label{3g_split}
\ee
where $\g_{\alpha\mu\nu}(q,r,p)$ represents the pole-free 
component, while $V_{\alpha\mu\nu}(q,r,p)$, 
whose origin is purely non-perturbative, 
contains 
all pole-related contributions. 
As we will discuss 
in detail in the next sections, the composition of $V_{\alpha\mu\nu}(q,r,p)$ 
is rather elaborate; however, for the purposes of 
creating a mass for  
the gluon propagator in the Landau gauge, only a minimal structure 
of $V_{\alpha\mu\nu}(q,r,p)$ is required, namely\footnote{In previous works \cite{Aguilar:2016vin,Aguilar:2017dco,Binosi:2017rwj,Aguilar:2021uwa,Aguilar:2022thg}, $\Rc1(q,r,p)$ has been denoted as $C_1(q,r,p)$.}
\be 
\label{eq:fatG} 
	V_{\alpha\mu\nu}(q,r,p) =  \frac{q_{\alpha}}{q^2} g_{\mu\nu} \Rc1(q,r,p) + \cdots \,, 
\ee
where all omitted terms drop out when $V_{\alpha\mu\nu}(q,r,p)$  
is inserted in diagrams $(a_1)$.

A detailed analysis reveals 
that~\cite{Aguilar:2016vin}  
\begin{equation}
	\Rc1(0,r,-r) = 0 \,; \label{V1_0} 
\end{equation}
therefore, the Taylor expansion of $\Rc1(q,r,p)$ around $q= 0$ yields
\be 
\label{eq:taylor_C}
	\lim_{q \to 0} \Rc1(q,r,p) = 2 (q\cdot r) \, \Cfat(r)
+ \, {\cal O}(q^2) \,,
\qquad \qquad 
\Cfat(r) := \left[ \frac{\partial \Rc1(q,r,p)}{\partial p^2} \right]_{q = 0}\,.
\ee
With the aid of \1eq{eq:taylor_C},
and after the extraction of the appropriate 
tensorial structure, 
the integral 
associated with the diagram $(a_1)$ yields 
\be 
m^2_{(a_1)} =  - \Delta_{(a_1)}^{-1}(0) = 3\lambda Z_3 \int_k k^2 \Delta^2(k)\Cfat(k) \,,\label{mass_origin}
\ee 
with
\be
\lambda := ig^2C_{\rm A}/2 \,,
\label{lambda}
\ee
where $C_\mathrm{A}$ is the Casimir eigenvalue of the adjoint representation [$N$ for SU$(N)$].
In the above formula, 
$Z_3$ stands for the 
renormalization constant of the three-gluon vertex, and we denote by 
\be\label{eq:int_measure}
\int_{k} := \frac{1}{(2\pi)^4} \int \!\!{\rm d}^4 k \,
\ee
the integration over virtual momenta; 
the use of a symmetry-preserving regularization scheme is implicitly assumed.

We next use standard rules (see eg~\cite{Aguilar:2021uwa}) to rewrite 
\1eq{mass_origin} in 
Euclidean space; note, in particular, that $m^2 = \Delta_{\srm E}(0)$.
Then, using hyperspherical coordinates, 
we obtain 
\be 
m^2_{(a_1)} = \frac{3\alpha_s C_{\rm A} Z_3}{8 \pi} \int_0^\infty \!\!\!\!dy\, y^2 \Delta_{\srm E}^2(y)\left\vert \Cfat_{\srm E}(y) \right\vert \,,\label{mass_euc}
\ee
with $\alpha_s := g^2/(4\pi)$ and $y := k_{\srm E}^2$.
Evidently, $m^2$ depends on the renormalization point, $\mu$; in particular, $m = 348$~MeV for $\mu = 4.3$~GeV~\cite{Aguilar:2021okw,Horak:2022aqx}\footnote{
A renormalization-group-invariant gluonic mass scale of about half the proton mass has been  obtained from the process-independent QCD effective charge~\cite{Binosi:2016nme,Cui:2019dwv}.}.

We emphasize that $\Cfat_{\srm E}(q)$, in addition to providing the gluon mass through \1eq{mass_euc} and its two-loop extension, plays a central role in this entire construction due to its dual nature. 
In particular: 

({\it i})  $\Cfat_{\srm E}(r)$ is 
the {\it BS amplitude} describing the formation of  gluon-gluon {\it colored} composite bound states;

\begin{figure}[t]
\includegraphics[width=0.47\linewidth]{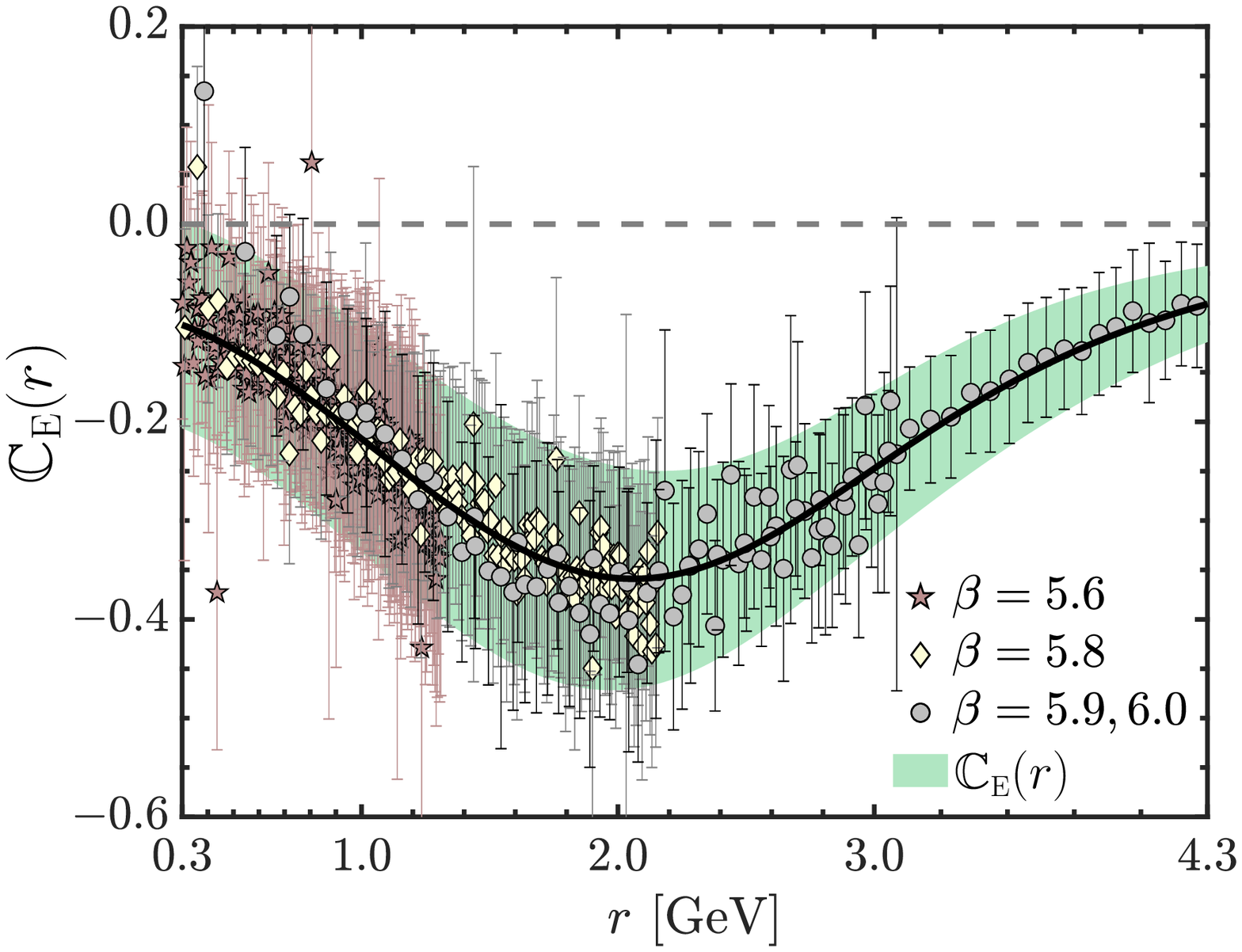}
\caption{The displacement function,  
$\Cfat_{\srm E}(r)$, obtained  from~\cite{Aguilar:2022thg,NarcisoFerreira:2023kak}. }
\label{fig:Cfat}
\end{figure}

({\it ii})
$\Cfat_{\srm E}(r)$ leads to 
a characteristic 
displacement of the WI satisfied by the 
pole-free part of the three-gluon vertex; 
for that reason, $\Cfat_{\srm E}(r)$ is 
called ``{\it displacement function}''. 
This predicted displacement has been  
confirmed by combining judiciously the results of several lattice simulations~\cite{Aguilar:2022thg,NarcisoFerreira:2023kak}; 
as shown in \fig{fig:Cfat}, 
the result for $\Cfat_{\srm E}(r)$ is clearly nonvanishing.

\section{Schwinger poles of the three-gluon vertex}\label{sec:background}

In this section we elaborate on the pole content of the 
three-gluon vertex, which arises as a consequence of the 
activation of the Schwinger mechanism. 
Our analysis relies on the bound-state interpretation 
of the poles associated with the Schwinger mechanism
(see \eg ~\mbox{\cite{Aguilar:2011xe,Aguilar:2017dco,Binosi:2017rwj,Aguilar:2016vin,
Aguilar:2021uwa,Aguilar:2022thg}}), making 
extensive use of the diagrammatic structure of the SDE of the three-gluon vertex.

\begin{figure}[t]
\includegraphics[width=0.8\linewidth]{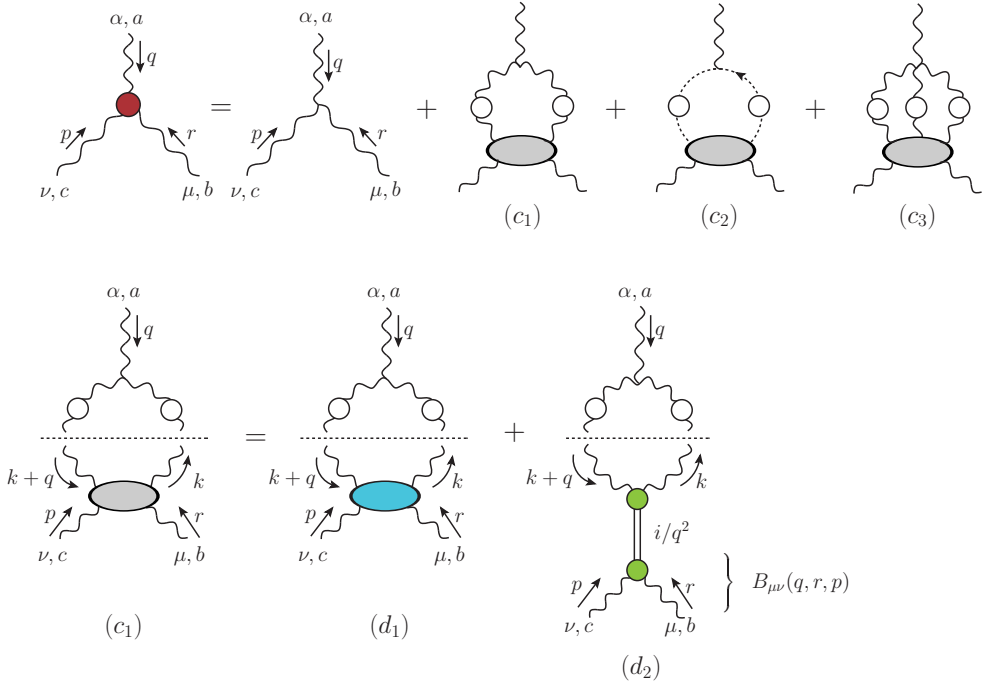}
\caption{Top: skeleton expansion of the three-gluon vertex.
The gray ellipses denote multi-particle kernels, which are 
one-particle irreducible with respect to the $q$-channel. Bottom: decomposition of the kernel of diagram $(c_1)$ into a term 
with no poles in the $q$ channel, denoted by $(d_1)$ (blue ellipse), and a term that 
contains a massless bound state, with propagator $i/q^2$, denoted by $(d_2)$.
For the purpose of clearer visualization 
of the various structures, 
the components of the four-gluon kernel are separated from the corresponding vertex graphs 
by dotted horizontal lines.}
\label{fig:4gkernel}
\end{figure}

The dynamics of $\fatg_{\alpha\mu\nu}(q,r,p)$
are determined by the SDE shown in panel $(A)$ of \fig{fig:4gkernel}. Following the standard way of writing the SDE of a vertex, 
a particular gluon leg of $\fatg_{\alpha\mu\nu}(q,r,p)$ 
is singled out (in this case the leg carrying momentum $q$), 
and is connected to the various multiparticle kernels 
through all 
elementary vertices of the theory. 
The remaining two legs (with momenta $r$ and $p$) are attached to the multiparticle kernels
through fully-dressed vertices.
Note that the full SDE  
is Bose-symmetric, albeit not manifestly so\footnote{Within the $n$PI effective action formalism, the resulting 
SDE for the three-gluon vertex 
is manifestly Bose-symmetric with respect to all of its three legs~\cite{Berges:2004pu,Carrington:2010qq,York:2012ib,Carrington:2013koa,Williams:2015cvx,Huber:2020keu}.};
in order to expose its Bose symmetry, 
the detailed skeleton expansion of the kernels must be 
taken into account. 

The seed of the Schwinger mechanism may be traced 
inside the four-particle kernel
appearing in the top panel of \fig{fig:4gkernel}.
It is triggered by 
the emergence of a 
colored scalar excitation, 
formed as a bound state of 
a pair of gluons, 
as shown pictorially  
in the bottom panel of \fig{fig:4gkernel};
note that the propagator of the composite scalar 
is given by $i\delta^{ab}/q^2$.
The resulting  
scalar-gluon-gluon interaction 
is described by the tensor 
denoted by $B_{\mu\nu}(q,r,p)$ in the bottom panel of \fig{fig:4gkernel}. 
The dynamics of $B_{\mu\nu}(q,r,p)$ 
is determined by solving 
the linear homogeneous 
BSE, which arises as the 
limit $q \to 0$ of the  
SDE for $\fatg_{\alpha\mu\nu}(q,r,p)$ is taken.
The nontrivial solution 
that one obtains corresponds
to the ``BS amplitude'' 
for the formation of a massless 
scalar out of two gluons. 
As explained in detail in~\cite{Papavassiliou:2022wrb}, the BS amplitude 
coincides, up to an overall scaling factor, with the 
displacement function $\Cfat(q)$.

\begin{figure}[t]
\includegraphics[width=0.8\linewidth]{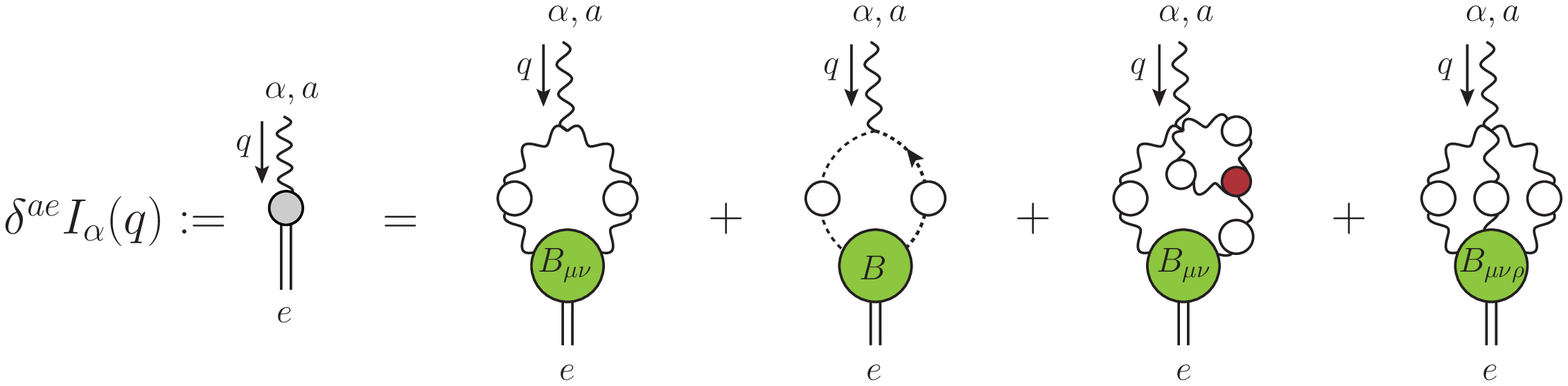}
\caption{Definition of the scalar-gluon transition amplitude, $I_\alpha(q)$.
The green circles represent 
the 
so-called ``proper vertex functions'' 
or ``bound state wave functions''~\cite{Jackiw:1973ha}. 
In particular, 
the $B_{\mu\nu}$,  
first introduced in 
\fig{fig:4gkernel}, 
describes the 
effective 
interaction between 
a composite scalar and two gluons, 
while 
$B$ and $B_{\mu\nu\rho}$ describe 
the interaction 
of a composite scalar with a 
ghost-antighost pair and 
three gluons, respectively.
}
\label{fig:I_def}
\end{figure}

When the upper part 
of the four-gluon kernel 
(legs with $k+q$ and $-k$)
is connected to the 
external gluon (with momentum $q$)
in order 
to form the three-gluon vertex, as shown in the bottom panel of \fig{fig:4gkernel}, the part that contains the 
composite scalar gives rise to the 
transition amplitude 
$I_{\alpha}(q)$, defined in \fig{fig:I_def}. Lorentz invariance imposes that 
$I_{\alpha}(q) = I(q) \, q_{\alpha}$, where $I(q)$ is a scalar function, whose 
role and properties have been discussed in detail in~\cite{Ibanez:2012zk,Papavassiliou:2022wrb}; note, in 
particular, the exact relation $m^2 = g^2 I^2(0)$.
As a consequence, 
the massless poles are   
{\it longitudinally coupled}~\mbox{\cite{Jackiw:1973tr,Cornwall:1973ts,Eichten:1974et,Poggio:1974qs,Smit:1974je}}, 
giving rise to tensorial structures of the general form 
$q_\alpha/q^2$, $r_{\mu}/r^2$, 
and $p_{\nu}/p^2$. Therefore, the pole part, $V_{\alpha\mu\nu}(q,r,p)$, 
satisfies the important relation 
\be 
P^{\alpha}_{\alpha'}(q) P^{\mu}_{\mu'}(r) P^{\nu}_{\nu'}(p) V_{\alpha\mu\nu}(q,r,p)  = 0 \,. \label{PPPV}
\ee

Note, in addition, that when
$V_{\alpha\mu\nu}(q,r,p)$ is  
contracted by two transverse projectors, 
only the poles in the uncontracted channel survive, {\it e.g.}\footnote{In the language of \1eq{Vbasis}, 
$P^{\mu}_{\mu'}(r) P^{\nu}_{\nu'}(p) V_{\alpha\mu\nu}(q,r,p) = \frac{q_\alpha}{q^2}P^{\mu}_{\mu'}(r) P^{\nu}_{\nu'}(p) \left[ \Rc1 g_{\mu\nu} + \Rc2 p_\mu r_\nu \right] $. },
\be 
P^{\mu}_{\mu'}(r) P^{\nu}_{\nu'}(p) V_{\alpha\mu\nu}(q,r,p) = \text{ only poles in }q^2 \,.
\label{PPV}
\ee

The nonlinear nature of the SDE 
makes $V_{\alpha\mu\nu}(q,r,p)$ 
contain mixed poles, of the type 
$q_\alpha r_{\mu}/q^2r^2$, etc. In the Landau gauge, these 
additional terms do not affect the 
gluon mass, which only depends on the residue of the single pole that coincides with the external momentum of the 
gluon SDE ($q$ in the conventions of \fig{fig:gluonself}).
Nonetheless, this type of pole 
is crucial for maintaining gauge invariance, by 
balancing properly the STI satisfied by $\g_{\alpha\mu\nu}(q,r,p)$.
In order to appreciate 
how such terms arise, 
we make the following two key observations.

\begin{figure}[t]
\includegraphics[width=1\linewidth]{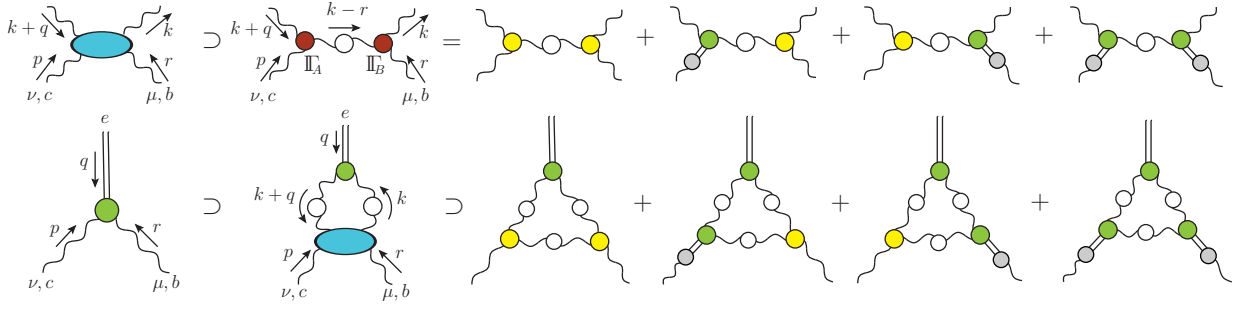}
\caption{Top: one-gluon exchange form of the blue kernel introduced in  \fig{fig:4gkernel}, and its 
subsequent decomposition into pole-free part (yellow vertices) and terms containing poles in $r$ and $p$. Bottom: decomposition of the scalar-gluon-gluon interaction, $B_{\mu\nu}(q,r,p)$, defined in of \fig{fig:4gkernel}, into pole-free and pole terms. }
\label{fig:4gkernel_Bose}
\end{figure}

({\it i})
To begin with,  
the part of the vertex with no poles 
in the channel $q$, 
contains poles in the other two ($r$ and $p$).
This is because the kernel associated with this part (blue ellipse in \fig{fig:4gkernel})
contains fully dressed vertices, 
as indicated 
schematically in the top panel of \fig{fig:4gkernel_Bose},
for the case of the ``one-gluon exchange'' approximation. 
Denoting by 
$\fatg_{\!\!\!\s A} := \fatg (p,k+q, r-k)$ 
and 
$\fatg_{\!\!\!\s B} := \fatg (r,k-r,-k)$,
as indicated in \fig{fig:4gkernel_Bose}, 
the contribution to the $(d_1)$ of \fig{fig:4gkernel} may be schematically 
written as 
\be 
(d_1) \sim \int_k \!\!\g^{(0)} \, \Delta \, \fatg_{\!\!\!\s A}\, \Delta 
\, \fatg_{\!\!\!\s B} \, \Delta 
\sim \int_k \!\!\g^{(0)} \, \Delta \, (\g_{\!\!\!\s A} + V_{\s A}) \, \Delta 
\, (\g_{\!\!\!\s B} + V_{\s B})\, \Delta \,,
\ee
where
\be 
\g^{(0)}_{\alpha\mu\nu}(q,r,p) = ( q - r )_\nu g_{\alpha\mu} + ( r - p )_\alpha g_{\mu\nu} + ( p - q )_\mu g_{\nu\alpha} \,, \label{3g_bare}
\ee
is the tree-level expression of the three-gluon vertex.

Then, using \1eq{3g_split}, 
and noting  
that the vertices 
$V_{\s A}$ and $V_{\s B}$ 
furnish 
poles only in the 
external momenta
$p$, and $r$, respectively, 
since poles in all other 
directions
are annihilated by the 
Landau gauge propagators 
[see \1eq{PPV}], 
one obtains  
\bea 
(d_1) \,\sim \,\underbrace{\int_k\! \g^{(0)} \! \Delta \, \g_{\!\!\!\s A} \,  \Delta 
\,\g_{\!\!\!\s B} \,  \Delta}_{\mbox{no pole}} 
\,+\,  \underbrace{\int_k \! \g^{(0)} \! \Delta \, V_{\!\s A} \, \Delta 
\,\g_{\!\!\!\s B} \, \Delta}_{\mbox{pole in} \,\; \displaystyle{p^2}}
\,+\, \underbrace{\int\! \g^{(0)} \! \Delta \, \g_{\!\!\!\s A}\,  \Delta\, V_{\!\s B}\, \Delta}_{\mbox{pole in} \,\; \displaystyle{r^2}}
\,+\,  \underbrace{ \int_k\! \g^{(0)}\!  \Delta  \,V_{\!\s A}  \,\Delta 
 \,V_{\!\s B}  \,\Delta}_{\mbox {poles in}\,\; \displaystyle{r^2}, \, \displaystyle{p^2}} \,. \nonumber\\
\label{parts}
\eea

({\it ii}) Furthermore, the same kernel appears 
in the part of the vertex 
describing the 
pole in the $q$-channel.
Thus, as indicated in the bottom panel of \fig{fig:4gkernel_Bose}, one obtains contributions of the 
type
\be 
(d_2) \sim \int \!\! V\, \Delta \, \fatg_{\!\!\!\s A}\, \Delta 
\, \fatg_{\!\!\!\s B} \, \Delta 
\sim \int \!\! V \, \Delta \, (\g_{\!\!\!\s A} + V_{\!\s A}) \, \Delta 
\, (\g_{\!\!\!\s B} + V_{\!\s B})\, \Delta \,,
\ee
giving rise to 
\bea 
(d_2) \,\sim \,\underbrace{\int\! V \! \Delta \, \g_{\!\!\!\s A} \,  \Delta 
\,\g_{\!\!\!\s B} \,  \Delta}_{\mbox {pole in}\,\; \displaystyle{q^2}} 
\,+\,  \underbrace{\int \! V \! \Delta \, V_{\!\s A} \, \Delta 
\,\g_{\!\!\!\s B} \, \Delta}_{\mbox {poles in}\,\; \displaystyle{q^2}, \, \displaystyle{p^2}}
\,+\, \underbrace{\int\! V \! \Delta \, \g_{\!\!\!\s A}\,  \Delta\, V_{\!\s B}\, \Delta}_{\mbox {poles in}\,\; \displaystyle{q^2}, \, \displaystyle{r^2}}
\,+\,  \underbrace{ \int\! V \!  \Delta  \,V_{\!\s A}  \,\Delta 
 \,V_{\!\s B}  \,\Delta}_{\mbox {poles in}\,\; \displaystyle{q^2}, \,\displaystyle{r^2}, \,\displaystyle{p^2}} \,. \nonumber\\
\label{parts2}
\eea 

\begin{figure}[t]
\includegraphics[width=1\linewidth]{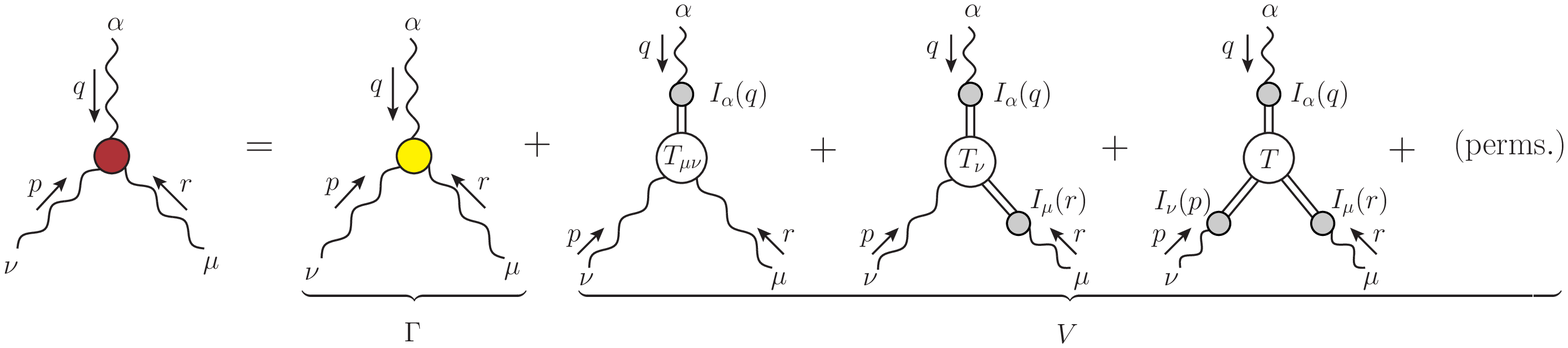}
\caption{ The general structure 
of the three-gluon vertex after the 
activation of the Schwinger 
mechanism. Note, in particular, 
that the term $V_{\alpha\mu\nu}(q,r,p)$ contains single poles, such as $q_\alpha/q^2$, as well as mixed poles of the forms $q_\alpha r_\mu/q^2 r^2$ and $q_\alpha r_\mu p_\nu/q^2 r^2 p^2$. The term  ``(perms)'' denotes the permutations of the external legs that lead to a Bose-symmetric $V_{\alpha\mu\nu}(q,r,p)$.}
\label{fig:3gpoles}
\end{figure}

The main conclusion of the analysis presented in this section is 
summarized in 
\fig{fig:3gpoles}, 
where  
\1eq{3g_split} is represented pictorially. 
In particular, the 
component $V_{\alpha\mu\nu}(q,r,p)$ is comprised 
by single poles, mixed double poles, and 
and mixed triple pole, depending on the 
number of gluon-scalar transition amplitudes (grey circles) 
contained in them. The
three types 
of effective amplitudes,  $T_{\mu\nu}(q,r,p)$, 
$T_{\mu}(q,r,p)$, and $T(q,r,p)$ (white circles)
are completely pole-free; 
see also \1eq{TV}.

We end this section with two final comments.

First, as we will 
demonstrate in Sec.~\ref{triple}, 
the triple mixed pole 
is not genuine; 
its strength 
is reduced due to  
requirements imposed 
by the self-consistency 
of the vertex STI, 
or, at the diagrammatic level, by virtue of 
\1eq{eq:taylor_C}.

Second, the validity of 
\1eq{PPPV}, which, in the bound-state formulation 
of the Schwinger mechanism arises naturally, 
guarantees 
that the lattice ``observables'' 
of the general form 
\be 
L(q,r,p) =  \frac{\lambda_{\alpha\mu\nu}(q,r,p)P_{\alpha\alpha'}(q)P_{\mu\mu'}(r)P_{\nu\nu'}(p)\fatg^{\alpha'\mu'\nu'}(q,r,p)}{\lambda_{\alpha\mu\nu}(q,r,p)P_{\alpha\alpha'}(q)P_{\mu\mu'}(r)P_{\nu\nu'}(p)\g_0^{\alpha'\mu'\nu'}(q,r,p)}  \,, \label{Lsg_def}
\ee 
where the $\lambda_{\alpha\mu\nu}(q,r,p)$ are appropriate projectors, 
are completely pole-free. 
Indeed, all lattice results obtained thus far
show no trace of pole divergences~\cite{Cucchieri:2006tf,Maas:2007uv,Cucchieri:2008qm,Athenodorou:2016oyh,Duarte:2016ieu,Boucaud:2017obn,Sternbeck:2017ntv,Vujinovic:2018nqc,Aguilar:2019uob,Aguilar:2021lke,Maas:2020zjp,Pinto-Gomez:2022brg,Pinto-Gomez:2023lbz,Boucaud:2018xup}.

\section{A purely longitudinal basis}\label{sec:basis}

In this section we introduce an appropriate basis for describing the special component  
$V^{\alpha\mu\nu}(q,r,p)$, which, due to the 
condition \1eq{PPPV}, is strictly longitudinal. 

As is well known, the most general Lorentz decomposition of the three-gluon vertex is comprised of 14 independent tensors. However, the strict longitudinality condition of \1eq{PPPV} imposes 4 constraints on the form factors of $V^{\alpha\mu\nu}(q,r,p)$. As a result, $V^{\alpha\mu\nu}(q,r,p)$ can be decomposed in a basis comprised of 10 tensors, 
denoted by $v_i^{\alpha\mu\nu} (q,r,p)$, 
accompanied by the associated form factors, 
denoted by $\Vc{i}(q,r,p)$, 
\ie 
\begin{align}
V^{\alpha\mu\nu}(q,r,p) =&
\sum_{i=1}^{10} \Vc{i}(q,r,p) \, 
v_i^{\alpha\mu\nu} (q,r,p)
\,,
\label{Vbasis_1}
\end{align}
where 
\be
\begin{tabular}{lll}
$v_1^{\alpha\mu\nu} = q^\alpha g^{\mu\nu}\,,$ & 
$v_2^{\alpha\mu\nu} = q^\alpha p^\mu r^\nu\,,$
&
$v_3^{\alpha\mu\nu} = r^\mu g^{\nu\alpha}\,, $\\
$v_4^{\alpha\mu\nu} = r^\mu q^\nu p^\alpha\,,$
&
$v_5^{\alpha\mu\nu}  = p^\nu g^{\alpha\mu}\,,$
&
$v_6^{\alpha\mu\nu} = p^\nu r^\alpha q^\mu\,,$
\\
$v_7^{\alpha\mu\nu} = q^\alpha r^\mu ( q - r )^\nu \,,$
\qquad\qquad &
$v_8^{\alpha\mu\nu} = r^\mu p^\nu ( r - p )^\alpha \,,$
\qquad\qquad &
$v_9^{\alpha\mu\nu} = p^\nu q^\alpha ( p - q )^\mu \,,$
\\
$v_{10}^{\alpha\mu\nu} = q^\alpha r^\mu p^\nu \,.$ & &
\end{tabular}
\label{vi_def}
\ee

Note that these tensors form 
three distinct groups, depending on the number of momenta to which they are longitudinal: 
the $v_i^{\alpha\mu\nu}$ with $i = 1, \ldots, 6$ 
are longitudinal to a single 
momentum, those with $i = 7, \ldots, 9$ to two, 
while $v_{10}^{\alpha\mu\nu}$ is longitudinal to all three momenta.

Now, following the bound state interpretation, each form factor $\Vc{i}$ can have massless poles in each of the channels to which the corresponding tensor, $v_i^{\alpha\mu\nu}$, is longitudinal. 
In particular, 
exhibiting the poles explicitly, we have 
\begin{align}
&\Vc1 = \frac{\Rc1}{q^2} \,; \qquad \; \Vc3 = \frac{\Rc3}{r^2}  \,; \qquad\;  \Vc5 = \frac{\Rc5}{p^2} \,; \qquad\;  \Vc7 = \frac{\Rc7}{q^2 r^2} \,; \qquad\;   \Vc9 = \frac{\Rc9}{p^2 q^2 } \,; \nonumber\\
&\Vc2 = \frac{\Rc2}{q^2} \,; \qquad \; \Vc4 = \frac{\Rc4}{r^2} \,; \qquad\;  \Vc6 = \frac{\Rc6}{p^2} \,; \qquad\;   \Vc8 = \frac{\Rc8}{r^2 p^2} \,; \qquad\;   \Vc{10} = \frac{\Rc{10}}{q^2 r^2 p^2} \,, 
\label{Vi_poles}
\end{align}
where the $\Rc{i} \equiv \Rc{i}(q,r,p)$ are 
regular functions, which, in the appropriate 
limits, capture the corresponding pole residues.

With the above definitions, $V_{\alpha\mu\nu}(q,r,p)$ can be recast in the form
\begin{align}
V_{\alpha\mu\nu}(q,r,p) &=
\frac{q_{\alpha}}{q^2}
\left(
g_{\mu\nu}\Rc1 + p_{\mu}r_{\nu}\Rc2
\right) \,\,
+ \,\,
\frac{r_{\mu}}{r^2}
\left(
g_{\alpha\nu}\Rc3 + q_{\nu}p_{\alpha}\Rc4
\right)\,\,
+ \,\,
\frac{p_{\nu}}{p^2}
\left(
g_{\alpha\mu} \Rc5 + r_{\alpha} q_{\mu}\Rc6
\right) \nonumber\\
& + \frac{q_{\alpha}r_{\mu}}{q^2 r^2}(q - r)_{\nu} \Rc7 \,\,
+ \,\, \frac{r_{\mu}p_{\nu}}{r^2 p^2}(r - p)_{\alpha} \Rc8 \,
+ \frac{p_{\nu}q_{\alpha}}{p^2 q^2}(p - q)_{\mu} \Rc9  \,\,
+ \,\, \frac{q_{\alpha} r_{\mu} p_{\nu}}{q^{2} r^{2} p^{2}}\,\Rc{10}
\,.
\label{Vbasis}
\end{align}

It is clear from the 
 diagrammatic representation of \fig{fig:3gpoles}, that $V_{\alpha\mu\nu}(q,r,p)$
is Bose-symmetric.
Consequently, and 
given that the color factor $f^{abc}$ 
has been factored out, we have that 
\be 
V_{\alpha\mu\nu}(q,r,p) = - V_{\mu\alpha\nu}(r,q,p) = - V_{\nu\mu\alpha}(p,r,q) \,. \label{VBose_explicit}
\ee
Then, from \2eqs{Vbasis}{VBose_explicit}
follows that the form factors 
$\Rc{i}(q,r,p)$ satisfy the following symmetry relations
\bea 
\Rc{1,2}(q,r,p) &= - \Rc{1,2}(q,p,r)\,, \qquad \quad  \Rc7(q,r,p)&= \Rc7(r,q,p) \,, \nonumber\\
\Rc{3,4}(q,r,p) &= - \Rc{3,4}(p,r,q)\,,  \qquad  \quad  \Rc8(q,r,p) &= \Rc8(q,p,r) \,, \nonumber\\
\Rc{5,6}(q,r,p) &=  -\Rc{5,6}(r,q,p)\,,  \qquad  \quad  \Rc9(q,r,p) &= \Rc9(p,r,q) \,,
\label{VBose}
\eea 
with $\Rc{10}(q,r,p)$ being totally anti-symmetric. In addition, some form factors are related to each other by the cyclic permutations of their arguments, namely 
\bea
\Rc{3,4}(q,r,p) &= \Rc{1,2}(r,p,q) \,, \qquad \quad   \Rc8(q,r,p) &= \Rc7(r,p,q) \,, \nonumber\\
\Rc{5,6}(q,r,p) &= \Rc{1,2}(p,q,r) \,, \qquad \quad   \Rc9(q,r,p) &= \Rc7(p,q,r) \,. 
\label{VBose_2}
\eea

In the limit $q\to 0$, we obtain from \1eq{VBose}  the 
relations 
\bea 
\Rc1(0,r,-r) &=& \Rc3(r, 0, -r) = \Rc5(r,-r,0) = 0 \,, \nonumber\\
\Rc2(0,r,-r) &=& \Rc4(r, 0, -r) = \Rc6(r,-r,0) = 0 \,, \nonumber\\
\Rc{10}(0,r,-r) &=& \Rc{10}(r,0,-r) = \Rc{10}(r,-r,0) = 0 \,. \label{q0_Bose}
\eea
The relations derived above will be employed in the analysis presented in the following 
sections. 

Finally, it is instructive to make contact between the form 
of the vertex 
$V_{\alpha\mu\nu}(q,p,r)$ 
given in \1eq{Vbasis}
and the pictorial 
representation of the 
same vertex,
depicted in \fig{fig:3gpoles}. 
In particular, 
the effective amplitudes 
$T_{\mu\nu}$, 
$T_{\mu}$, and $T$ 
may be expressed in terms
of the form factors $V_i$
through the 
direct matching of the various 
tensorial structures, 
namely 
\begin{align} 
I(q) T_{\mu\nu}(q,r,p)=&\, - i \left[ g_{\mu\nu}\Rc{1}(q,r,p) + p_\mu r_\nu \Rc2(q,r,p) \right] \,, \nonumber\\
I(q) I(r) T_\nu(q,r,p) =&\, (r - q)_\nu \Rc7(q,r,p) \,, \nonumber \\
I(q) I(r) I(p) T(q,r,p)=&\, i \Rc{10}(q,r,p) \,. \label{TV}
\end{align}
%


We conclude this discussion with some remarks regarding the basis given by \1eq{vi_def}. 
The 14 tensors required for the full description of $\fatg^{\alpha\mu\nu}(q,r,p)$ 
may be obtained  
by supplementing the $v_i^{\alpha\mu\nu}$ of \1eq{vi_def} with 4 totally transverse tensors, say ${\overline t}_i^{\,\alpha\mu\nu}$, such that
$
q_\alpha{\overline t}_i^{\,\alpha\mu\nu} = r_\mu{\overline t}_i^{\,\alpha\mu\nu} = p_\nu{\overline t}_i^{\,\alpha\mu\nu} = 0 \,.$
For example, 
one could use the $t_i^{\alpha\mu\nu}$ given in Eq.~(3.6) of \cite{Aguilar:2019jsj}, corresponding to the transverse part of the Ball-Chiu (BC) basis~\cite{Ball:1980ax}. However, the resulting basis, \mbox{${v_i^{\alpha\mu\nu}}\cup{t_j^{\alpha\mu\nu}}$}, introduces spurious divergences in certain form factors of the pole-free part, thus 
being unsuitable for many applications. 
Furthermore, as explained in~\cite{Aguilar:2019jsj}, the BC basis is inconvenient for  
the description of 
$V^{\alpha\mu\nu}(q,r,p)$, because the 10 non-transverse tensors (the $\ell_i^{\alpha\mu\nu}$ in Eq.~(3.4) of \cite{Aguilar:2019jsj}) are not longitudinal\footnote{In~\cite{Ball:1980ax}, the  $\ell_i^{\alpha\mu\nu}$ span 
the part of the vertex that saturates the STI, 
which was denominated  
``longitudinal'', in contradistinction to the ``transverse'' (automatically conserved) component.
The confusion caused by the fact that the $\ell_i^{\alpha\mu\nu}$ are not longitudinal, in the 
sense explained above, may be avoided by 
using the term ``non-transverse'' instead.}, in the sense 
that $P_{\alpha^\prime\alpha}(q)P_{\mu^\prime\mu}(r)P_{\nu^\prime\nu}(p)\ell_i^{\alpha\mu\nu}\neq 0$. As a result, in the BC basis, the transverse components of $V^{\alpha\mu\nu}(q,r,p)$ acquire poles as well, which combine in complicated ways with the non-transverse ones to yield a strictly longitudinally coupled $V^{\alpha\mu\nu}(q,r,p)$. 
The basis of~\cite{Eichmann:2014xya} 
appears to suffer from the same shortcoming. 
Thus, it is preferable to decompose the pole-free and pole parts in different bases, such as the BC for $\g^{\alpha\mu\nu}(q,r,p)$ and \1eq{vi_def} for $V^{\alpha\mu\nu}(q,r,p)$.

\section{ Mixed poles from the Slavnov-Taylor identity}\label{sec:3g_const}

In this section, we turn our attention to the STI satisfied by the full vertex $\fatg_{\alpha\mu\nu}(q,r,p)$. 
As we will show in detail, when the 
gluon propagator is finite at the origin (massive), the STI
imposes an extended  
pole content on
the three-gluon vertex.
Specifically, 
the only way to achieve self-consistency  
is by introducing 
mixed poles in $V_{\alpha\mu\nu}(q,r,p)$; the form factors associated 
with these poles must satisfy 
strict constraints, which 
preclude their vanishing.

We emphasize that the central assumption underlying this analysis 
is that
both the BRST symmetry\footnote{ We employ 
the \emph{standard} BRST symmetry
of QCD, to be distinguished from 
the modified BRST symmetry of the refined Gribov-Zwanziger
action~\cite{Dudal:2008sp,Capri:2015ixa}.}
and the associated STIs 
remain intact 
when the gluon acquires a mass
through the action of the 
Schwinger mechanism. This assumption is strongly corroborated 
by the STI-driven extraction of the 
$\Cfat (q)$ using lattice inputs
~\cite{Aguilar:2021uwa,Aguilar:2022thg,NarcisoFerreira:2023kak,Ferreira:2023fva}; for 
a variety of related discussions
and approaches,  
see~\cite{Kugo:1979gm,Alkofer:2000wg,Fischer:2008uz,Alkofer:2010cwc,Alkofer:2011pe,Cyrol:2016tym,Pawlowski:2022oyq}, and references therein.

\subsection{ Abelian STI with a hard mass}\label{subsec:3gab}

To fix the ideas, let us consider first the simplified situation
where the three-gluon vertex satisfies the Abelian STI given by
\be 
q^\alpha \fatg_{\alpha\mu\nu}(q,r,p) = 
P_{\mu\nu}(p)\Delta^{-1}(p) - 
P_{\mu\nu}(r)\Delta^{-1}(r) \,. \label{STIint}
\ee
Moreover, let the gluon propagator be given by the tree-level form, \ie \mbox{$\Delta^{-1}(q) \to q^2 - m^2$}, corresponding to 
a simple massive propagator
in Minkowski space.

Then, after substitution, 
the STI becomes 
\be 
q^\alpha\fatg_{\alpha\mu\nu}(q,r,p) = g_{\mu\nu}(p^2 - r^2) + r_\mu r_\nu - p_\mu p_\nu + m^2\left( \frac{p_\mu p_\nu}{p^2} - \frac{r_\mu r_\nu}{r^2} \right) \,. \label{STIint_explicit}
\ee
Evidently, the form factors 
associated with 
the tensor structures $p_\mu p_\nu$ and $-r_\mu r_\nu$ on 
the r.h.s. of \1eq{STIint_explicit} 
contain poles 
in $p^2$ and $r^2$, respectively, 
whose residue is $m^2$. 
In fact, 
these tensor structures are 
longitudinal to the uncontracted legs of the vertex, \ie those carrying momenta $r$ and $p$.

Hence, the self-consistency of \1eq{STIint_explicit} requires that $\fatg_{\alpha\mu\nu}(q,r,p)$ 
should 
contain longitudinally coupled poles of the form $r_\mu/r^2$ and $p_\nu/p^2$. Evidently, from the cyclic permutations of \1eq{STIint_explicit} (equivalently, from Bose symmetry),  $\fatg_{\alpha\mu\nu}(q,r,p)$ must also contain massless poles longitudinally coupled to $q_\alpha$, \ie of the form $q_\alpha/q^2$. Thus, the STI implies that $\fatg_{\alpha\mu\nu}(q,r,p)$ must assume the special form of \1eq{3g_split}, with a nonzero pole part, $V_{\alpha\mu\nu}(q,r,p)$.

At this point, let us assume that the pole-free part of the vertex, $\g_{\alpha\mu\nu}(q,r,p)$, reduces to the tree-level form given in \1eq{3g_bare}, \ie $\g_{\alpha\mu\nu}(q,r,p) \to \g^{(0)}_{\alpha\mu\nu}(q,r,p)$, such that, 
from 
\1eq{3g_split}, $\fatg_{\alpha\mu\nu}(q,r,p)=\g^{(0)}_{\alpha\mu\nu}(q,r,p) 
+V_{\alpha\mu\nu}(q,r,p)$.
Then, \1eq{STIint_explicit} can be recast into an STI for $V_{\alpha\mu\nu}(q,r,p)$, namely
\be 
q^\alpha V_{\alpha\mu\nu}(q,r,p) = m^2 \left( \frac{p_\mu p_\nu}{p^2} - \frac{r_\mu r_\nu}{r^2} \right) \,, \label{V_STI}
\ee
together with its cyclic permutations.

Next, assume that $V_{\alpha\mu\nu}(q,r,p)$ satisfies the longitudinality condition of \1eq{PPPV}, as required by  
both the Schwinger mechanism and lattice QCD.
Expanding out the transverse projectors in that equation, and using \1eq{V_STI} and its permutations, one straightforwardly obtains
\be 
V_{\alpha\mu\nu}(q,r,p) = \frac{m^2}{2}\left[ \frac{q_\alpha r_\mu}{q^2 r^2}(q-r)_\nu + \frac{r_\mu p_\nu}{r^2p^2 }(r-p)_\alpha + \frac{p_\nu q_\alpha}{p^2 q^2}(p-q)_\mu  \right]\,, \label{V_Cornwall}
\ee
which shows that $V_{\alpha\mu\nu}(q,r,p)$ must contain mixed double poles, with residues proportional to the gluon mass.
Note that this result amounts to the constant mass limit of the \emph{Ansatz} given in \cite{Ibanez:2012zk}, constructed therein for a momentum-dependent mass, $m^2(q)$. Furthermore, the combination $\g^{(0)}_{\alpha\mu\nu}(q,r,p)+V_{\alpha\mu\nu}(q,r,p)$, with $V_{\alpha\mu\nu}(q,r,p)$ given by \1eq{V_Cornwall} reproduces the effective three-gluon vertex of Cornwall~\cite{Cornwall:1985bg,Cornwall:2010upa}.
Lastly, comparing \2eqs{Vbasis}{V_Cornwall} we read off the expressions for the form factors
\be 
\Rc7(q,r,p) = \Rc8(q,r,p) = \Rc9(q,r,p) = \frac{m^2}{2} \,,
\ee
with all other $\Rc{i}$ vanishing in this simple case.

If the form of the pole-free part is not known, as is generally the case, the complete momentum dependence of $V_{\alpha\mu\nu}(q,r,p)$ cannot be determined. Nevertheless, the values of the form factors $\Rc{i}(q,r,p)$ of \1eq{Vbasis} at zero momenta can be obtained unequivocally from the STI. In particular, in the toy model of \1eq{STIint_explicit}
\be 
\Rc9(q) = m^2/2 \,, \label{V9_0_Abelian}
\ee
independently of the exact form of $\g_{\alpha\mu\nu}$, where we use \1eq{VBose} and define
\be 
\Rc9(q) := \Rc9(q,-q,0) = \Rc9(0,q,-q) \,.
\ee 
Evidently, the same result holds for $\Rc8(q,-q,0) = \Rc8(q,-q,0) = \Rc7(q,0,-q) = \Rc7(0,q,-q)$.

\subsection{ General case: mixed poles and the residue function}\label{subsec:3gnab}

Having fixed the 
general ideas, we now turn 
to the full form of the STI, and 
demonstrate how to obtain from it expressions for 
the $\Rc{i}$ when one of the momenta vanishes. 

The STI is 
given by~\cite{Marciano:1977su}
\be 
q^\alpha \fatg_{\alpha\mu\nu}(q,r,p) = F(q)\left[ \Delta^{-1}(p) P^\alpha_\nu(p) H_{ \alpha \mu }(p,q,r) - \Delta^{-1}(r) P^\alpha_\mu(r) H_{\alpha\nu}(r,q,p) \right] \,; \label{STI}
\ee
the cases 
$r^\mu \fatg_{\alpha\mu\nu}(q,r,p)$
and 
$p^\nu \fatg_{\alpha\mu\nu}(q,r,p)$
are obtained from \1eq{STI} through permutations of 
the appropriate momenta and indices.
In the above equation,  
$F(q)$ is the ghost dressing function, defined in terms of the ghost propagator $D^{ab}(q) = i\delta^{ab}D(q)$ by $D(q) = F(q)/q^2$, while $H_{ \mu \nu }(r,q,p)$ represents the ghost-gluon scattering kernel, with $r,\,q,\,p$ denoting the momenta of the anti-ghost, ghost, and gluon, respectively.

The most general Lorentz structure of $H_{ \mu \nu }(r,q,p)$ is given by~\cite{Aguilar:2018csq} 
\be 
H_{\mu\nu}(r,q,p) = g_{\nu\mu} A_1 + r_\mu r_\nu A_2 + p_\mu p_\nu {\mathbb A}_3 + p_\mu r_\nu A_4 + r_\mu p_\nu {\mathbb A}_5 \,, \label{Htens}
\ee
where $A_i\equiv A_i(r,q,p)$ and ${\mathbb A}_i\equiv {\mathbb A}_i(r,q,p)$; the 
use of distinct notation for the third and fifth form factors will become clear in what follows.
At tree level, $A_1^{(0)} = 1$, while all other form factors vanish. Note that since in \1eq{STI} the 
$H_{\mu\nu}(r,q,p)$ is contracted by transverse projectors, 
only the form factors $A_1$, $A_4$ and ${\mathbb A}_3$ contribute to the STI. 

At this point, it is crucial to recognize 
that the Schwinger mechanism
induces poles not only to the vertex 
$\fatg_{\alpha\mu\nu}(q,r,p)$ but also to the 
ghost-gluon kernel $H_{\mu\nu}(r,q,p)$. 
In particular, the poles are longitudinally 
coupled, carrying the momentum and Lorentz index
of the incoming gluon leg. 
Therefore, they are contained in 
the form factors ${\mathbb A}_{3,5}(r,q,p)$, which assume the general form 
\be 
{\mathbb A}_{3,5}(r,q,p) = \frac{A_{3,5}^\p(r,q,p)}{p^2} + A_{3,5}(r,q,p) \,, \label{Hpoles}
\ee
where $A_{3,5}(r,q,p)$ denotes the pole-free part.

To determine the residues of the poles required by the STI, we begin by decomposing both sides of \1eq{STI} in the same basis and equating coefficients of independent tensor structures. Since the tensors appearing in the STI have two free Lorentz indices and two independent momenta, they can all be decomposed in the same basis employed for $H_{\mu\nu}(r,q,p)$ in \1eq{Htens}. In particular, the 
contracted pole-free part may be written as
\be 
q^\alpha\g_{\alpha\mu\nu}(q,r,p) = \qG1 g_{\mu\nu} + \qG2 r_\mu r_\nu + \qG3 p_\mu p_\nu + \qG4 p_\mu r_\nu + \qG5 r_\mu p_\nu \,, \label{qGamma_expansion} 
\ee
with $\qG{i}\equiv \qG{i}(q,r,p)$. At tree level, $\qG1^{(0)} = p^2 - r^2$, $\qG2^{(0)} = 1$, $\qG3^{(0)} = -1$, and \mbox{$\qG4^{(0)} = \qG5^{(0)} = 0$}. Note that, from Bose symmetry, $\qG1$, $\qG4$ and $\qG5$ must be anti-symmetric under the exchange of $r\leftrightarrow p$, such that
\be 
\qG1(0,r,-r) = \qG4(0,r,-r) = \qG5(0,r,-r) = 0 \,. \label{Gi_0}
\ee 
Then, since $\g_{\alpha\mu\nu}(q,r,p)$ is pole-free, its contraction with $q^\alpha$ vanishes when $q = 0$, such that \2eqs{qGamma_expansion}{Gi_0} imply also that
\begin{align}
\qG2(0,r,-r) = - \qG3(0,r,-r) \,. \label{G23_0}
\end{align}

As for the pole part, after contracting \1eq{Vbasis} with $q^\alpha$, we obtain
\be 
q^\alpha V_{\alpha\mu\nu}(q,r,p) = \Rc1 g_{\mu\nu} + \frac{\qV2}{r^2} r_\mu r_\nu + \frac{\qV3}{p^2} p_\mu p_\nu + \Rc2 p_\mu r_\nu + \frac{\qV5}{r^2 p^2} r_\mu p_\nu \,, \label{qV_tens}
\ee 
with the $\qV{i} \equiv \qV{i}(q,r,p)$ given by 
\begin{align}
\qV2 =& - \Rc3 - ( p\cdot q ) \Rc4  - 2\Rc7  \,, \qquad
\qV3 = - \Rc5 - ( r\cdot q ) \Rc6  + 2\Rc9  \,, \nonumber\\
\qV5 =& \Rc{10} + (p^2 - r^2) \Rc8 - p^2 \left[ \Rc3 + (q\cdot p) \Rc4 + \Rc7 \right] - r^2 \left[ \Rc5 + ( q\cdot r ) \Rc6 - \Rc9 \right] \,. \label{Vs_2basis}
\end{align}

As in the previous subsection, 
we now isolate the tensor structures $r_\mu r_\nu$ and $p_\mu p_\nu$ on both sides of \1eq{STI}; equating their coefficients yields
\begin{align}
\qG2 =&\, \frac{1}{r^2}\left\lbrace F(q)\left\lbrace \Delta^{-1}(r) \left[ A_1(r,q,p) + ( p\cdot r )A_4(r,q,p) \right] + r^2 \Delta^{-1}(p) {\mathbb A}_3(p,q,r) \right\rbrace \!-\!\qV2 \right\rbrace\,, \nonumber \\
\qG3 =&\, - \frac{1}{p^2}\left\lbrace F(q)\left\lbrace p^2 \Delta^{-1}(r) {\mathbb A}_3(r,q,p) + \Delta^{-1}(p) \left[ A_1(p,q,r) + ( p\cdot r )A_4(p,q,r) \right] \right\rbrace + \qV3 \right\rbrace\,. 
\label{pmupnu}
\end{align}
Since $\qG3$ is pole-free by definition, in the limit $p\to 0$, 
the term $1/p^2$ must be 
canceled by the content of the 
curly bracket in \1eq{pmupnu}.
Thus, what 
appears to be a pole in $p^2$ 
must be converted into an evitable singularity. 
The condition for this to occur 
is given by 
\be 
\Rc9(q) = \frac{F(q)}{2}\left[ m^2 A_1(q) - \Delta^{-1}(q) A_3^\p(q) \right] \,, \label{V9_0}
\ee
where $m^2 = - \Delta^{-1}(0)$, in Minkowski space, and we used \1eq{Hpoles} and defined
\be 
A_1(q) := A_1(0,q,-q)\,, \qquad A_3^\p(q) := A_3^\p(q,-q,0) \,. \label{Aiq_def}
\ee
Note that setting 
the ghost-sector Green's functions 
in \1eq{V9_0} to their tree level
expressions 
in \1eq{V9_0}, \ie $F\to 1$, $A_1\to 1$ and $\mathbb{A}_3\to0$, leads to \1eq{V9_0_Abelian}.

Similarly, the requirement that the $\qG2$ of \1eq{pmupnu} be pole-free at $r = 0$ yields a relation identical to \1eq{V9_0}, but with $\Rc9(q)$ substituted by $\Rc7(0,q,-q)$. This last result follows also from Bose symmetry, according to \1eq{VBose}. For the same reason, \1eq{V9_0} also holds with the left-hand side substituted by any one of $\Rc7(0,q,-q)$, $\Rc8(q,-q,0)$, and $ \Rc8(q,0,-q)$.

Returning to \1eq{V9_0}, 
it is clear that the 
only way for $\Rc9$ to vanish 
identically for all $q$
(\ie for $V_{\alpha\mu\nu}$ not to contain 
the associated mixed pole) is 
for the r.h.s. to also vanish; 
however, at least when $q=0$, 
this cannot happen
in the Landau gauge.
Indeed, as was demonstrated in 
~\cite{Aguilar:2020yni}, in this gauge, 
\be 
A_1(0) = {\widetilde Z}_1 \,, \qquad A_3^\p(0) = 0 \,, \label{A3p_0}
\ee
where ${\widetilde Z}_1$ is the renormalization constant of the ghost-gluon vertex, which is \emph{finite} by virtue of Taylor's theorem~\cite{Taylor:1971ff}. Hence, at the origin, \1eq{V9_0} reduces to
\be 
\Rc9(0) = \frac{1}{2}{\widetilde Z}_1F(0) m^2 \,. \label{V9_to_Delta}
\ee 
Consequently, just as in the toy model of \1eq{STIint_explicit}, the self-consistency of the full STI 
in the presence of an infrared finite 
gluon propagator requires the appearance of a pole associated with the form factor $\Rc9$.

The function $\Rc9(q)$,
associated with the 
mixed pole $1/q^2 p^2$
will be particularly important in the analysis that follows. 
To understand its nature,
consider a function of two variables, $x$ and $y$ 
of the form 
$f(x,y) = g(x,y)/xy$, 
with $g(x,0) \neq 0$ 
and $g(0,y)\neq 0$, such that 
$f(x,y)$
has simple poles as 
$x \to 0$ and  $y\to 0$. 
In particular, if we take 
$y\to 0$, the residue of this 
pole is a function of $x$, given by $r(x) = g(x,0)/x$.
In fact, if we subsequently take $x \to 0$, $g(0,0)\neq 0$ is the residue of the function 
$r(x)$. 
Evidently, in this analogy, $g(x,0)$ 
plays the role of $\Rc9(q)$; 
in what follows, we will 
refer to  $\Rc9(q)$ as the \emph{``residue function''}. 

We conclude this discussion by pointing out that the tensor structures $g_{\mu\nu}$, $r_\nu p_\mu$ and $r_\mu p_\nu$ of \1eq{STI}, associated with the pole-free form factors $\qG{1,4,5}$, lead to constraints on the behavior of the remaining form factors, $\Rc{i}$, which are absent from the 
simplified result 
of \1eq{V_Cornwall}. These additional relations, however, constrain certain derivatives of the $\Rc{i}$, rather than the values of the form factors themselves. Indeed, the constraint obtained from $g_{\mu\nu}$ is equivalent to the so-called ``Ward Identity displacement'', which has been analyzed in detail in recent works~\cite{Aguilar:2021uwa,Aguilar:2022thg,Papavassiliou:2022wrb,Ferreira:2023fva}. On the other hand, the $r_\mu p_\nu$ structure leads to a constraint on the form factor $\Rc{10}$, which 
amounts to a drastic reduction 
of the triple pole associated with 
it; the detailed demonstration of this 
point is given in Section~\ref{triple}.

\section{Residue function from the Schwinger-Dyson equation}\label{V9SDE}

The special relation given in \1eq{V9_0} implies that, in the presence of an infrared finite gluon propagator, 
the appearance of mixed poles 
in the three-gluon vertex is an inevitable requirement 
of the STI. In this section we explore this same relation  
from the point of view of the SDE satisfied by the 
three-gluon vertex. 
Specifically, we will show that, when the 
dynamical structures 
imposed by the activation of the Schwinger mechanism are duly taken into account, 
a truncated form of the vertex SDE leads 
to an approximate version of \1eq{V9_0}.

\subsection{General considerations}\label{V9SDEa}

In order to obtain 
from the vertex SDE 
the relation satisfied 
by the 
residue function $\Rc9(q)$ in \1eq{V9_0}, 
we follow 
the same procedure employed in its 
derivation from the STI: 
({\it i}) we begin by contracting the vertex SDE by $q_\alpha$; ({\it ii}) then, we isolate the tensor structure $p_\mu p_\nu$ from the result, which yields  $\qV{3}/p^2 + \qG{3}$ [recall \2eqs{qGamma_expansion}{qV_tens}]; ({\it iii}) finally, we multiply by $p^2$ and take the limit $p = 0$, where $\qV{3}(q,r,p)\to 2\Rc9(q)$ [see \1eq{Vs_2basis}].

\begin{figure}[ht]
\includegraphics[width=1\linewidth]{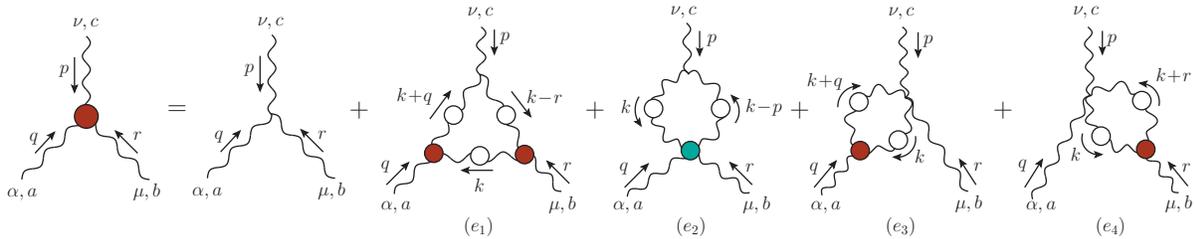}
\caption{The 
  diagrams contributing to the truncated SDE for 
three-gluon vertex that we employ; ghost and two-loop diagrams have been omitted. The swordfish diagrams $(e_{2,3,4})$ 
carry a symmetry factor of $1/2$.}
\label{fig:V9diags}
\end{figure}

To streamline the application of this procedure, it is convenient to set up the vertex SDE with tree-level vertices in the leg carrying momentum $p$, as in \fig{fig:V9diags}, rather than the version shown in \fig{fig:gluonself}. With this choice, the contraction of the SDE by $q^\alpha$ triggers inside the diagrams the STIs for the fully dressed vertices,
with an incoming $q$-leg. These STIs, in turn, simplify the identification of certain pole contributions stemming from the four-gluon vertex, as we will see shortly. 

We emphasize that 
the SDE given in  \fig{fig:V9diags} is truncated, 
by keeping only ``one-loop 
dressed diagrams'' containing 
gluons and the massless composite  
excitations associated 
with the Schwinger mechanism. 
Note, in particular, the absence of contributions originating from the ghost loop 
denoted by $(c_2)$ in 
\fig{fig:4gkernel}, and that 
the only representatives 
from graph $(c_3)$ are 
diagrams $(e_3)$ and $(e_4)$.
Given this truncation, 
we do not expect to reproduce 
\1eq{V9_0} in its entirety; 
in particular, it is reasonable to 
expect that the term 
$m^2$  will be approximated 
by its one-loop dressed gluonic 
expression, $m^2_{(a_1)}$,  
given in \1eq{mass_origin}. 
As we will see in what follows, this is indeed what happens.

Recalling the diagrammatic 
analysis presented in 
Sec.~\ref{sec:background},
it is relatively 
straightforward to establish 
that diagrams $(e_1)$, 
$(e_3)$, and $(e_4)$
contain poles 
in the $q$- and $r$-channels, 
but not in the $p$-channel,  
which is relevant for 
the derivation of \1eq{V9_0}.

Instead, $(e_2)$ possesses a pole 
in the $p$-channel, as may be 
deduced by means of two 
different (but ultimately equivalent) arguments, both 
related to the nature of the 
fully-dressed four-gluon vertex~\cite{Pascual:1980yu,Papavassiliou:1992ia,Hashimoto:1994ct,Driesen:1998xc,Kellermann:2008iw,Ahmadiniaz:2013rla,Gracey:2014ola,Cyrol:2014kca,Binosi:2014kka,Eichmann:2015nra,Gracey:2017yfi}, 
$\fatg_{\alpha\mu\delta\tau}^{abts}$.

The first argument is based on 
the observation 
that 
the special diagram $(d_2)$ 
of \fig{fig:4gkernel}
is part 
of $\fatg_{\alpha\mu\delta\tau}^{abts}$; 
evidently, since the 
vertex SDE is now written 
with respect to the $p$-leg, 
the appropriate replacement 
(\eg $q \to p$)
must be carried out in $(d_2)$,
which acquires thusly a massless 
bound state propagator $i/p^2$.

The second is by noticing 
that the STI satisfied by 
$\fatg_{\alpha\mu\delta\tau}^{abts}$
generates 
naturally a pole in the 
$p$-channel, provided that the  
poles of the 
three-gluon vertices appearing in its r.h.s. are properly included. 
Specifically, the STI reads 
\cite{Baker:1976vz,Kim:1979ep,Binosi:2009qm}
\begin{align}
q^\alpha\fatg_{\alpha\mu\delta\tau}^{abts}(q,r,p-k,k) &= F(q) \big[ f^{tad} f^{dbs}H^\gamma_{\hphantom{\gamma}\delta}(k+r,q,p-k)\fatg_{\gamma\mu\tau}(-k-r,r,k) \nonumber\\ 
& +  f^{sad}f^{dbt}H^\gamma_{\hphantom{\gamma}\tau}(-k-q,q,k)\fatg_{\gamma\mu\delta}(k+q,r,p-k) \nonumber\\ 
& + f^{bad}f^{dst}H^\gamma_{\hphantom{\gamma}\mu}(p,q,r)\underbrace{\fatg_{\gamma\tau\delta}(-p,k,p-k)}_{\text{Contains }\frac{p_\gamma}{p^2}} \big] + \ldots \,, \label{4gSTI}
\end{align} 
where the ellipsis denotes terms involving a ghost-ghost-gluon-gluon scattering kernel. Since this kernel has no tree-level value~\cite{Kim:1979ep,Binosi:2009qm}, we expect it to be subleading and omit it from our treatment.
Evidently, the full three-gluon vertices appearing on the r.h.s. of \1eq{4gSTI} enter  
with their entire  pole content. 
In particular, the underbrace in that equation highlights the explicit appearance of a $p_\gamma/p^2$ term, originating from the vertex $\fatg_{\gamma\tau\delta}(-p,k,p-k)$.

It is clear that the two arguments presented above are interlinked: the r.h.s. of the STI in  \1eq{4gSTI} contains a pole 
in $1/p^2$ because diagram $(d_2)$ 
is part of the four-gluon vertex, whose 
contraction by $q^\alpha$ appears on the l.h.s.
This accurate balance evidences once again 
the harmonious interplay 
between symmetry and dynamics.

\subsection{The derivation}\label{V9SDEab} 

Armed with these observations, we 
now proceed to the derivation of 
\1eq{V9_0},  
following the three main steps, 
({\it i})-({\it iii}),
mentioned above.

To that end, we start from the complete expression for $(e_2)$,
\begin{align} 
(e_2)_{\alpha\mu\nu}^{abc} &= - \frac{ig^2 Z_3}{2}\int_k\Delta^{\rho\delta}(k-p)\Delta^{\sigma\tau}(k)\g^{(0)}_{\nu\sigma\rho}(-p,k,p-k) f^{cst}\fatg_{\alpha\mu\delta\tau}^{abts}(q,r,p-k,k) \label{d2} \,,
\end{align}
where we have factored out a $g$. 

Then, following step ({\it i}),  
we contract \1eq{d2} with $q^\alpha$, thus triggering 
the STI of \1eq{4gSTI}.
Then, \1eq{d2} yields
\begin{align} 
q^\alpha (e_2)_{\alpha\mu\nu} =&\, - \frac{\lambda Z_3 F(q) }{p^2}\int_k\Delta^{\rho\delta}(k-p)\Delta^{\sigma\tau}(k)\g^{(0)}_{\nu\sigma\rho}(-p,k,p-k) p^\gamma H_{\gamma\mu}(p,q,r) \nonumber\\
\times&\, \left[ V_1(-p,k,p-k) g_{\delta\tau} + V_2(-p,k,p-k)p_{\delta}p_{\tau}\right] + \ldots \,, \label{qd2}
\end{align}
where the color structure $f^{abc}$ 
has been canceled out from both sides, and the ellipsis denotes terms that do not contain $1/p^2$ poles, and thus cannot contribute to $\Rc9(q)$.

Next, we evaluate the term $p^\gamma H_{\gamma\mu}(p,q,r)$ appearing in \1eq{qd2} using the well-known STI~\cite{Marciano:1977su}
\be
p^\gamma H_{\gamma\mu}(p,q,r) = \fatg_\mu(p,q,r) \,, \label{H_to_Gamma}
\ee
where $\fatg^{abc}_\mu(p,q,r) = - g f^{abc}\fatg_\mu(p,q,r)$ is the ghost-gluon vertex, whose most general Lorentz decomposition reads~\cite{Aguilar:2009nf,Aguilar:2018csq,Aguilar:2021okw}
\be 
\fatg_\mu(p,q,r) = p_\mu B_1(p,q,r) + r_\mu \mathbb{B}_2(p,q,r) \,.
\ee
Then, it follows from \1eq{H_to_Gamma} that
\begin{align} 
p^\gamma H_{\gamma\mu}(p,q,r) =&\, p_\mu B_1(p,q,r) + r_\mu \mathbb{B}_2(p,q,r)  \,, \label{pH}
\end{align}
while \1eq{Htens} allows us to write the $B_i$ in terms of the $A_i$ and $\mathbb{A}_i$ as~\cite{Aguilar:2009nf,Aguilar:2018csq}
\begin{align} 
B_1(p,q,r) =&\, A_1(p,q,r) + p^2 A_2(p,q,r) + (p\cdot r)A_4(p,q,r) \,, \nonumber\\
\mathbb{B}_2(p,q,r) =&\, (p\cdot r) \mathbb{A}_3(p,q,r) + p^2 \mathbb{A}_5(p,q,r) \,. \label{B1_B2}
\end{align}
Note that, since $\mathbb{A}_i(p,q,r)$ 
displays a pole when 
$r = 0$, so does the $\mathbb{B}_2(p,q,r)$; the pole amplitude associated with the ghost-gluon vertex has been studied in detail in~\cite{Aguilar:2017dco,Aguilar:2021uwa,Eichmann:2021zuv}. 

Now, we proceed to step ({\it ii}). Clearly, only the term $p_\mu B_1(p,q,r)$ of \1eq{pH} can contribute to the tensor structure $p_\mu p_\nu$, once inserted in \1eq{qd2} for $q^\alpha (e_2)_{\alpha\mu\nu}$. Hence, we can write 
\begin{align} 
q^\alpha (e_2)_{\alpha\mu\nu} =&\, - \frac{\lambda Z_3 F(q) B_1(p,q,r) p_\mu}{p^2}\int_k\Delta^{\rho\delta}(k-p)\Delta^{\sigma\tau}(k)\g^{(0)}_{\nu\sigma\rho}(-p,k,p-k)  \nonumber\\
\times&\, \left[ V_1(-p,k,p-k) g_{\delta\tau} + V_2(-p,k,p-k)p_{\delta}p_{\tau}\right] + \ldots \,, \label{qd2_H}
\end{align}
where the ellipsis now denotes terms that cannot contribute to $\Rc9(q)$ because they do not contain either a $1/p^2$ or a $p_\mu$.

In anticipation of the fact that we will take $p\to 0$ at the end of the calculation, we can already consider $p$ to be small. In this case, we can use into \1eq{qd2_H} the Taylor expansion given in \1eq{eq:taylor_C}, and its analog with $\Rc1$ substituted by $\Rc2$. Then, one sees that the term $\Rc2(-p,k,p-k)$ cannot contribute to $\Rc9(q)$, since it is two orders higher in $p$ than $\Rc1(-p,k,p-k)$. Hence, we obtain explicitly
\be 
q^\alpha (e_2)_{\alpha\mu\nu} = 3 \lambda  Z_3 F(q) B_1(p,q,r) \left(\frac{p_\mu p_\nu}{p^2} \right) \int_k k^2 \Delta^2(k) \Cfat(k) + \ldots \,, \label{qd2_Taylor}
\ee
with the ellipsis now including terms that are higher-order in the Taylor expansion around $p = 0$.

At this point, recalling \2eqs{qV_tens}{Vs_2basis}, the scalar coefficient of $p_\mu p_\nu$ in \1eq{qd2_Taylor} yields a contribution to $\qV{3}/p^2 + \qG{3}$, namely
\be 
\qV{3}^{(e_2)}(p,q,r)/p^2 + \qG{3}^{(e_2)}(p,q,r) = \frac{1}{p^2}F(q)B_1(p,q,r)\left[ 3\lambda  Z_3\int_k k^2 \Delta^2(k) \Cfat(k) \right] + \ldots \,, \label{V9_SDE_last_step}
\ee
where the superscript ``$(e_2)$'' emphasizes that the above expression contains only the contribution from diagram $(e_2)$.

Lastly, we perform step ({\it iii}), \ie multiply \1eq{V9_SDE_last_step} by $p^2$ and set $p = 0$. In doing so, we note from \1eq{B1_B2} that $B_1(0,q,-q) = A_1(q)$, while \2eqs{q0_Bose}{Vs_2basis} imply that $\qV{3}^{(e_2)}(0,q,-q) = 2\Rc9^{(e_2)}(q)$. Furthermore, since $(e_2)$ is the only diagram of \fig{fig:V9diags} that contributes to $\Rc9(q)$, we have \mbox{$\Rc9(q) = \Rc9^{(e_2)}(q)$}, such that
\be 
\Rc9(q) = F(q)A_1(q)\left[ \frac{3\lambda  Z_3}{2}\int_k k^2 \Delta^2(k) \Cfat(k) \right] = \frac{1}{2}F(q)A_1(q) m^2_{(a_1)} \,,
\ee
where we used \1eq{mass_origin} to obtain the last equality.

Therefore, the SDE of \fig{fig:V9diags} satisfies an approximate form of \1eq{V9_0}, where only the term containing $A_3^\p(q)$ in that equation is absent. This term could arise in the full SDE either from the diagrams that we omitted in \fig{fig:V9diags}, or from the ghost-ghost-gluon-gluon kernel that we dropped in \1eq{4gSTI};  its proper restoration requires a detailed treatment  
that goes beyond the scope of the present work.

Finally, we point out that, at 
$q=0$, the SDE result for $\Rc9(0)$ satisfies the STI requirement of \1eq{V9_to_Delta} \emph{exactly}, by virtue of \1eq{A3p_0}.

\section{Computing the Residue function}\label{num}

We next turn to the numerical determination of the residue function, $\Rc9(q)$, from \1eq{V9_0}. To this end, we first transform \1eq{V9_0} to Euclidean space, to obtain
\be 
\Rc9(q) = \frac{F(q)}{2}\left[ m^2 A_1(q) + \Delta^{-1}(q) A_3^\p(q) \right] \,. \label{V9_0_euc}
\ee
As we will explain below, 
for the determination of $A_3^\p(q)$ 
we will make use of the displacement function $\Cfat(q)$, shown in \fig{fig:Cfat}. Since 
in~\cite{Aguilar:2022thg,NarcisoFerreira:2023kak}
the $\Cfat(q)$ has been computed 
 in the so-called ``asymmetric MOM scheme''
~\cite{Athenodorou:2016oyh,Boucaud:2017obn,Aguilar:2020yni,Aguilar:2021okw}, 
 with $\mu=4.3$ GeV, 
 the same renormalization prescription will be employed in what follows. Note that, in this scheme, the finite 
 ghost-gluon renormalization constant appearing in \2eqs{A3p_0}{V9_to_Delta} is given by ${\widetilde Z}_1 = 0.9333$~\cite{Aguilar:2022thg,Ferreira:2023fva}.

\begin{figure}[ht]
\includegraphics[width=0.47\linewidth]{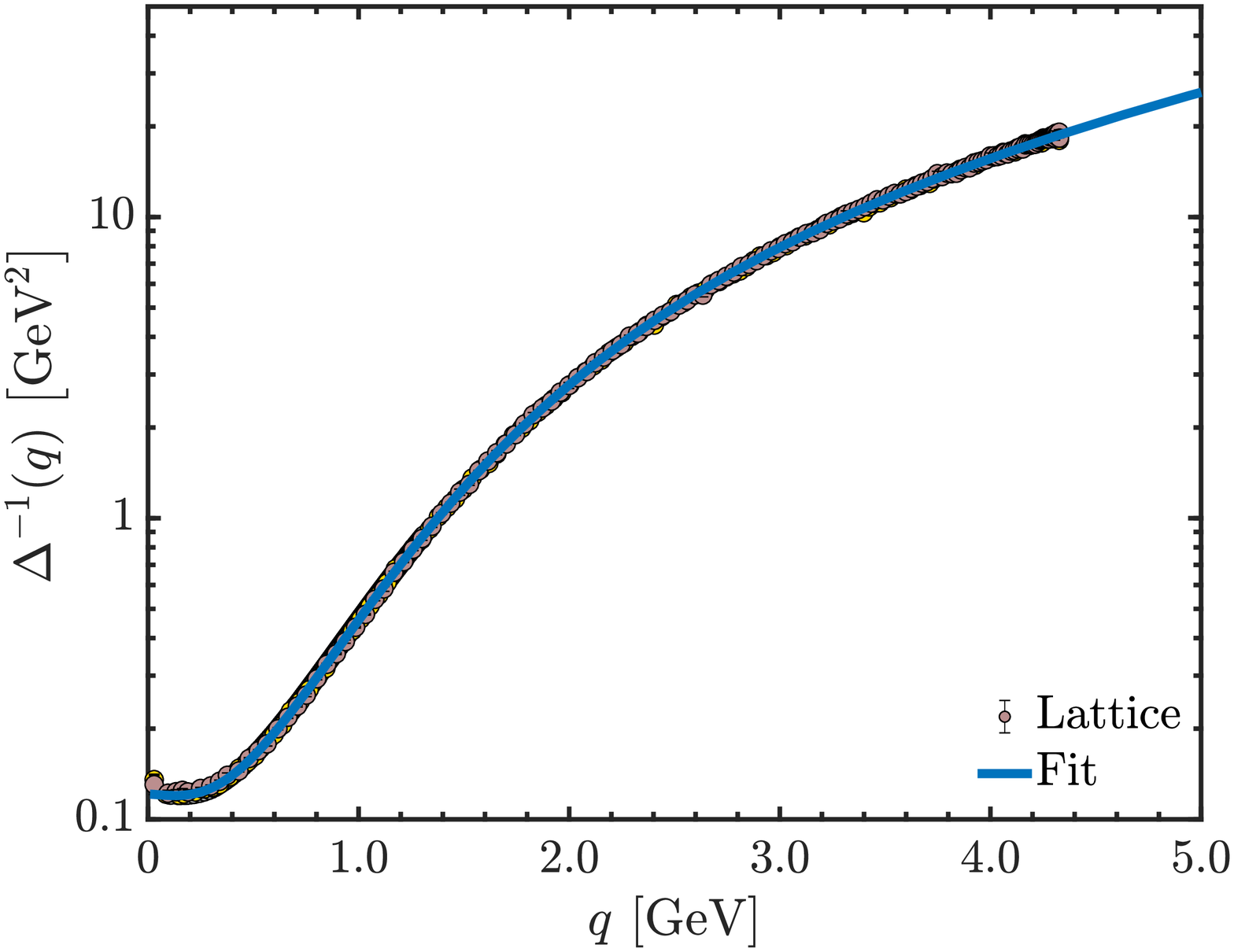} \hfil \includegraphics[width=0.47\linewidth]{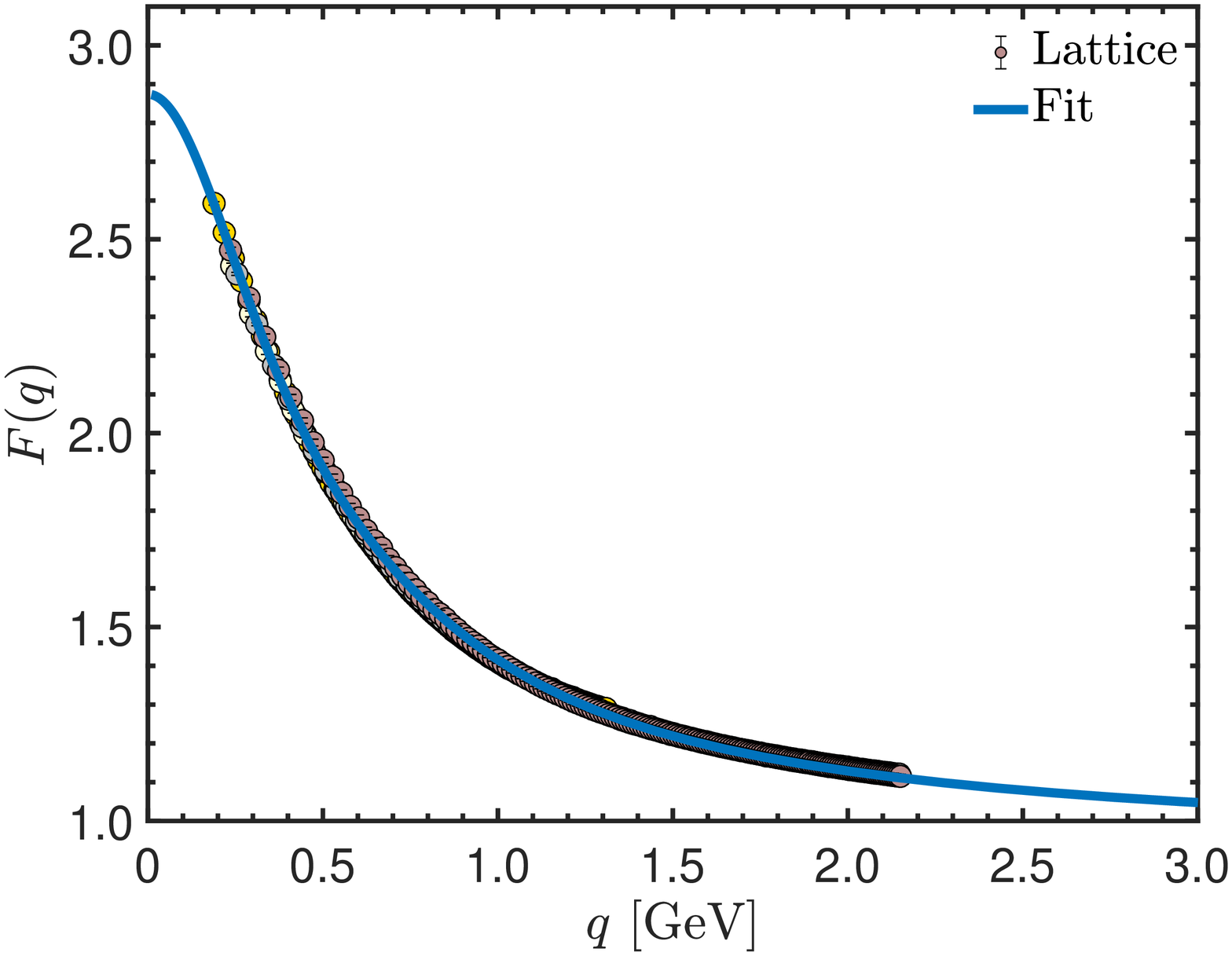} 
\caption{ Lattice data (points) of~\cite{Aguilar:2021okw} for $\Delta^{-1}(q)$ (left) and $F(q)$ (right), together with their corresponding fits (blue solid lines), given by Eqs.~(C6) and (C11) of~\cite{Aguilar:2021uwa}, respectively.}
\label{fig:inputs}
\end{figure}

Then, for the $F(q)$ and $\Delta(q)$ appearing in \1eq{V9_0_euc} we use physically motivated fits to lattice data of \cite{Aguilar:2021okw}, given by Eqs.~(C6) and (C11) of  \cite{Aguilar:2021uwa}, respectively. These fits are shown as continuous blue lines in \fig{fig:inputs}, where they are compared to the lattice data of \cite{Aguilar:2021okw} (points). Note that the value of $\Delta^{-1}(0) = 0.121$~GeV$^{-2}$ corresponding to this fit leads to the previously mentioned value of $m = 348$~MeV for $\mu = 4.3$~GeV. Moreover, the fitting functions for both $\Delta(q)$ and $F(q)$ were constructed in such a way that they reproduce the respective one-loop resummed anomalous dimensions.

\begin{figure}[ht]
\includegraphics[width=0.47\linewidth]{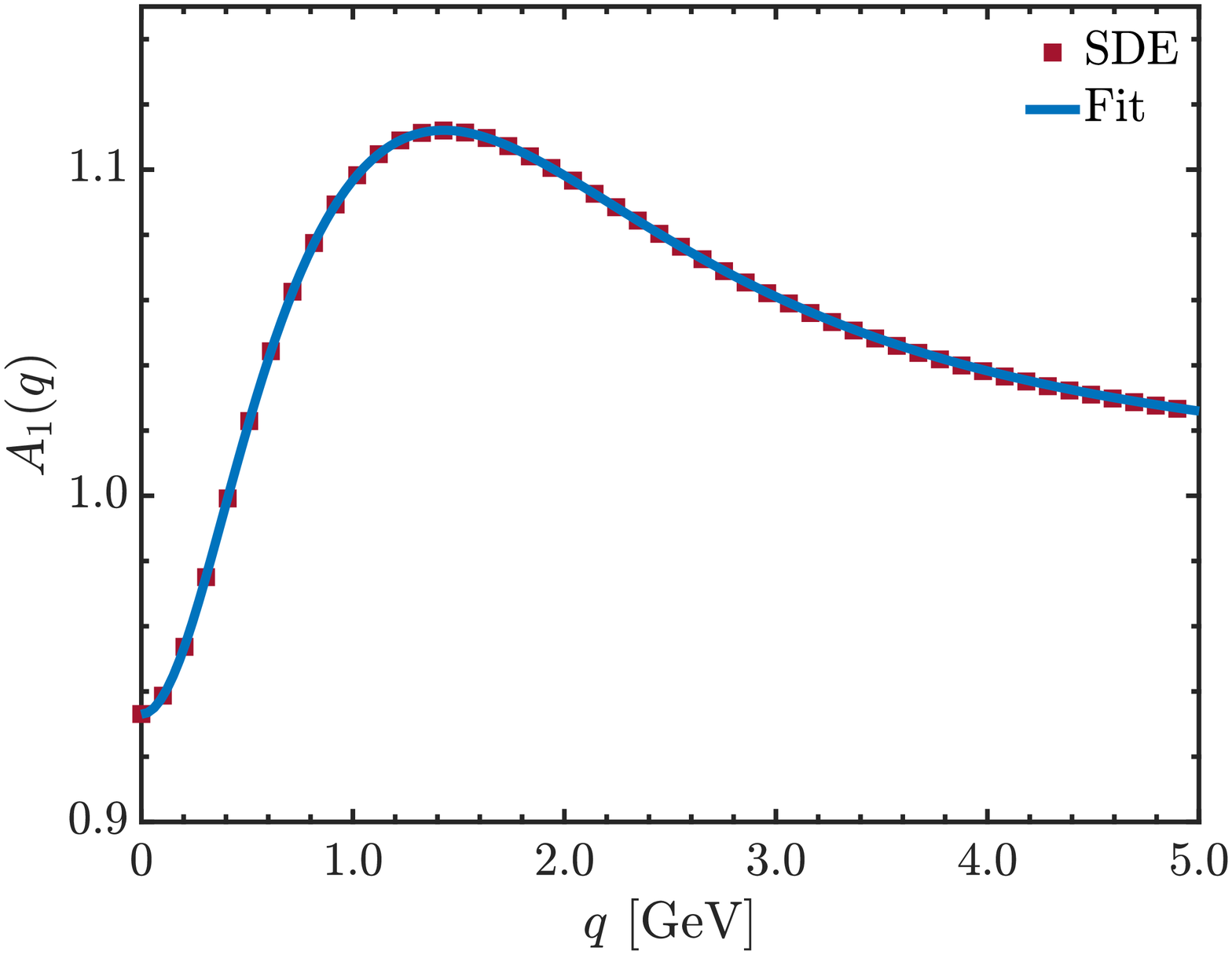} \hfil \includegraphics[width=0.47\linewidth]{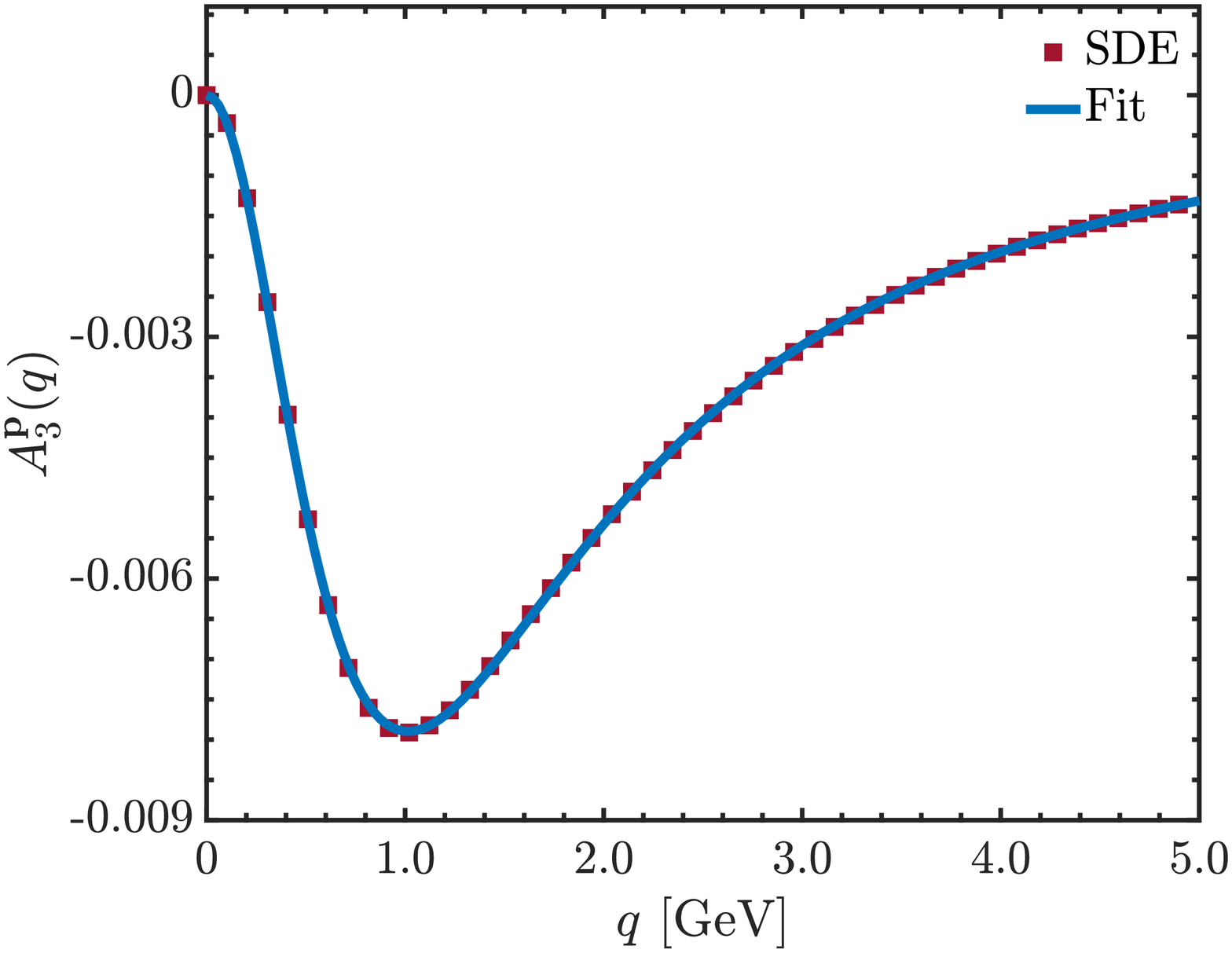}
\caption{ Left: The form factor $A_1(q)$ of the 
ghost-gluon kernel, in the 
soft-antighost limit, taken from~\cite{Aguilar:2018csq}. Right: Pole amplitude $A_3^p(q)$ of the ghost-gluon kernel, computed from the truncated SDE of \1eq{A3_eq}. }
\label{fig:Ai}
\end{figure}

For the  tree-level (classical) form factor $A_1(q)$ of the ghost-gluon kernel, we employ the result of the SDE analysis 
of~\cite{Aguilar:2018csq}\footnote{In~\cite{Aguilar:2018csq} the Taylor scheme~\cite{vonSmekal:2009ae,Boucaud:2008gn,Boucaud:2011eh,Zafeiropoulos:2019flq}
was employed; 
the conversion to the 
asymmetric scheme proceeds through the relation  
$A_1^{\rm asym} = {\widetilde Z} A_1^{\rm Taylor}$.}; 
the result is shown as red squares in the left panel of \fig{fig:Ai}, and is seen to deviate by at most $11\%$, at $q = 1.43$~GeV, from the tree-level value, $A_1^{(0)}(q) = 1$.

Then, the only unknown ingredient in \1eq{V9_0_euc} 
is the ghost-gluon pole term $A_3^\p(q)$, which may be 
computed as follows. 
We start with the one-loop dressed truncation of the SDE describing the ghost-gluon kernel, $H_{\mu\nu}(r,q,p)$, see, \eg Fig.~3 of \cite{Aguilar:2018csq}. From this SDE, we derive a dynamical equation for $\mathbb{A}_3(r,q,p)$, using the projector ${\cal T}_3^{\mu\nu}$, given in Eqs.~(3.7) and (3.8) of \cite{Aguilar:2018csq}. Then, recalling \1eq{Hpoles}, we obtain $A_3^\p(q)$ from the equation for $\mathbb{A}_3(r,q,p)$ by multiplying it by $p^2$ and taking the limit $p\to 0$. This procedure furnishes a \emph{linear} integral equation for $A_3^\p(q)$, which has the form
\be 
A_3^\p(q) = \int_k {\cal K}_1(k,q) A_3^\p(k) + \int_k {\cal K}_2(k,q)\Cfat(k) \,, \label{A3_eq}
\ee
where we notice the appearance of $\Cfat(q)$,  and the kernels ${\cal K}_i(k,q)$ are comprised by combinations of $\Delta$, $F$, and kinematic factors.

Then, we solve \1eq{A3_eq} numerically through the Nystr\"om method~\cite{Press:1992zz}, employing for $\Cfat(q)$ the result of~\cite{Aguilar:2022thg,NarcisoFerreira:2023kak}, shown in \fig{fig:Cfat}. Through this procedure, we obtain the result shown as red squares in the right panel of \fig{fig:Ai}.

For convenience, we provide fits for the functions $A_1(q)$ and $A_3^\p(q)$, which, in conjunction with the fits for $\Delta(q)$ and $F(q)$, allow $\Rc9(q)$ to be computed most expeditiously. Specifically, both $A_1(q)$ and $A_3^\p(q)$ can be accurately fitted by the low-degree rational functions
\be 
A_1(q) = {\widetilde Z}_1\left[ 1 + R_1(q) \right] \,, \qquad A_3^\p(q) = {\widetilde Z}_1 R_2(q) \,, \label{Ai_fits}
\ee
where
\be 
R_1(q) = \frac{ q^2/a_1 + \left( q^2/a_2 \right)^2 + \left( q^2/a_3 \right)^3  }{ 1 
+ q^2/b_1 + \left( q^2/b_2 \right)^2 + \left( q^2/b_3 \right)^3 } \,, \qquad R_2(q) = \frac{ q^2/c_1   }{ 1 
+ q^2/d_1 + \left( q^2/d_2 \right)^2 } \,,
\ee
with fitting parameters given by $a_1 = 
 1.71$~GeV$^2$, $a_2 = 2.68$~GeV$^2$, $a_3 = 4.51$~GeV$^2$, \mbox{$b_1 = 0.410$~GeV$^2$}, $b_2 = 1.30$~GeV$^2$, $b_3 = 1.89$~GeV$^2$, $c_1 = - 27.3$~GeV$^2$, $d_1 = 0.419$~GeV$^2$ and $d_2 = 1.03$~GeV$^2$.

We emphasize that the fits in \1eq{Ai_fits} preserve certain limits of the original functions, $A_1$ and $A_3^\p$. First, at the origin, the fits for $A_1(q)$ and  $A_3^\p(q)$ satisfy \1eq{A3p_0}. Next, a one-loop calculation reveals that, at large values of the momentum, $A_1(q)$ saturates to a constant~\cite{Aguilar:2018csq}; in addition, 
the numerical SDE result indicates that, 
in the same kinematic limit, 
$A_3^\p \sim 1/q^2$. It is straightforward to verify that these ultraviolet features are 
correctly captured
by the fits given by \1eq{Ai_fits}.

\begin{figure}[ht]
\includegraphics[width=0.47\linewidth]{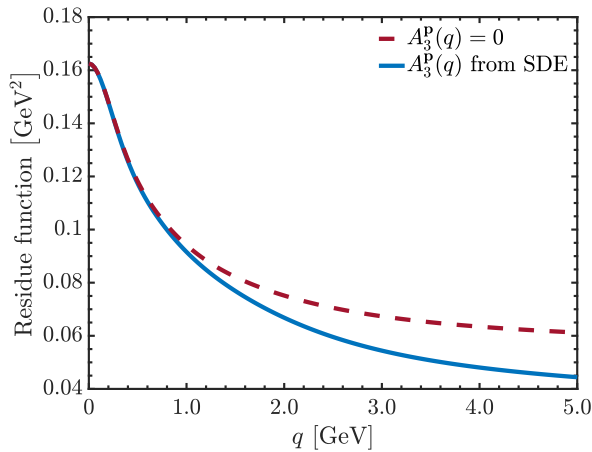}
\caption{ Residue function, $\Rc9(q)$, computed from \1eq{V9_0}, using the $A_3^\p(q)$ shown in \fig{fig:Ai} (blue continuous), compared to the result obtained if $A_3^\p(q)$ is set to zero (red dashed line). }
\label{fig:V9}
\end{figure}

Using the above ingredients in \1eq{V9_0_euc}, we obtain the $\Rc9(q)$ shown as a solid blue curve in \fig{fig:V9}. Comparing this result to that for $F(q)$, shown in the right panel of \fig{fig:inputs}, it is clear that the shape of $\Rc9(q)$ is dominated by the ghost dressing function in \1eq{V9_0_euc}.

Next, we test the effect of $A_3^\p$ on $\Rc9$. To this end, we set $A_3^\p(q)$ to zero in \1eq{V9_0_euc}, in which case we obtain the result shown as a red dashed line in \fig{fig:V9}. Note that the latter becomes equal to the full result (blue continuous) at $q = 0$, by virtue of \1eq{A3p_0}, but differs significantly from it for $q > 1$~GeV. 

To understand this difference, let us note that although $A_3^\p(q)$ itself is small in comparison to the corresponding pole of the three-gluon vertex, $\Cfat(q)$ [cf.~\fig{fig:Cfat} and~\fig{fig:Ai}], or even the saturation value of $\Rc9(q)$, it appears multiplied by $\Delta^{-1}(q)$. Since the latter increases rapidly in the ultraviolet, the product $A_3^\p(q)\Delta^{-1}(q)$ can contribute a considerable amount to $\Rc9(q)$ at large $q$, as indeed is observed.

\section{Absence of mixed triple pole \label{triple}}

In this section, we analyze the infrared behavior of the form factor 
$\Rc{10}(q,r,p)$, which is accompanied by a denominator $q^2 r^2 p^2$ in \1eq{Vbasis}. As  such, at first sight, one expects that this term should act as a triple mixed pole.

Consider, for instance, taking all momenta to zero by first taking $p \to 0$, which implies also $r = -q$, and then taking $q \to 0$. In this case, if $\Rc{10}(0,0,0)$ were nonvanishing, one would have
\be 
\lim_{q,p\to 0} \frac{ \Rc{10}(q,r,p) }{q^2r^2 p^2} = \frac{\Rc{10}(0,0,0)}{q^4 p^2} \,, \label{V10_q0_r0_naive}
\ee
where we use the shorthand notation
\be 
\lim_{q,p\to 0} := \lim_{q\to 0}\lim_{p\to 0} \,.
\ee

However, as we will demonstrate in Subsection~\ref{subsec:V10_STI},  the STI of \1eq{STI} requires $\Rc{10}(q,r,p)$ to vanish in this limit as 
\be 
\lim_{q,p\to 0}\Rc{10}(q,r,p) = 2(q\cdot p ) q^2 f(0) \,,\label{V10_q0_r0_Mink}
\ee
where $f(q)$ is some pole-free function at $q = 0$.
Consequently, in Euclidean space, 
\be 
\lim_{q,p\to 0}\frac{\Rc{10}(q,r,p)}{q^2 r^2 p^2} = \lim_{q,p\to 0} \frac{2\cos \theta}{|q| |p|} f(0) \,, \label{V10_q0_r0}
\ee
where $q\cdot p = |q||p|\cos\theta$ and $|q|$ denotes the magnitude of the Euclidean momentum $q$.

Moreover, an approximate SDE analysis  presented 
in Subsection~\ref{subsec:V10_dyn} shows  
that this particular requirement is 
enforced by the Schwinger mechanism, 
and especially due to the validity of \1eq{eq:taylor_C}.

Note that the divergence in \1eq{V10_q0_r0} is in fact weaker than that associated with the form factors $\Rc7$, $\Rc8$, and $\Rc9$, namely
\be 
\lim_{q,p\to 0} \frac{\Rc9(q,r,p)}{q^2p^2} = \lim_{q,p\to 0}\frac{{\widetilde Z}_1 F(0)m^2}{2 q^2 p^2}
\,,
\ee
where we used \1eq{V9_to_Delta}.

\subsection{Demonstration from the STI}\label{subsec:V10_STI}

For the derivation  of \1eq{V10_q0_r0_Mink} from the STI of \1eq{STI}, we begin by noting that \1eq{q0_Bose}
already implies that 
\be 
\lim_{p \to 0} \Rc{10}(q,r,p) = 2(p\cdot q) \Rc{10}^\prime(q) \,, \qquad \Rc{10}^\prime(q) := \left[ \frac{\partial \Rc{10}(q,r,p)}{\partial r^2} \right]_{p = 0} \,. \label{V10_Taylor}
\ee
Then, to obtain \1eq{V10_q0_r0_Mink} we need to show that
\be 
\lim_{q \to 0}\Rc{10}^\prime(q) = q^2 f(0) \,. \label{V10_goal}
\ee

In order to constrain $\Rc{10}$ from the STI, let us first note that this form factor appears in \1eq{STI}
through the combination $\qV5$, defined in \2eqs{qV_tens}{Vs_2basis}. Then, isolating its respective tensor structure, $r_\mu p_\nu$,
and equating the corresponding coefficients on each side of \1eq{STI}, we obtain 
\be 
\qG5 = \frac{1}{r^2 p^2} \left\lbrace ( p\cdot r) F(q) \left[ p^2  \Delta^{-1}(r) {\mathbb A}_3(r,q,p) - r^2 \Delta^{-1}(p) {\mathbb A}_3(p,q,r) \right] - \qV5 \right\rbrace\,, \label{rmupnu}
\ee
where we note the term $r^2 p^2$ in the denominator. Since $\qG5$ is pole-free, the r.h.s of \1eq{rmupnu} must be an evitable singularity at $p = 0$ and $r = 0$, which implies that the term in curly brackets must vanish sufficiently fast in those limits.

Then, since $\qG5$ is antisymmetric under the exchange or $r$ and $p$, it suffices to consider the $p = 0$ limit of \1eq{rmupnu}; hence, we expand the term in brackets around $p = 0$. 

Using the Bose symmetry relations of \2eqs{VBose}{VBose_2}, it is straightforward to show that the zeroth order term vanishes. However, the linear term yields the nontrivial constraint
\be 
\qV5^\prime(q) = - \frac{F(q)}{2}\left\lbrace m^2\left[ q^2 A_3(0,q,-q) + A_3^\p(0,q,-q) \right] + \Delta^{-1}(q)A_3^\p(q) \right\rbrace\,, \label{V10_step_0}
\ee
where we emphasize that $A_3^\p(q)$ [see \1eq{Aiq_def}] corresponds to a kinematic limit (soft-gluon)  different from $A_3^\p(0,q,-q)$ (soft-antighost), and define
\be
\qV5^\prime(q) := \left[ \frac{\partial V_5(q,r,p)}{\partial r^2} \right]_{p = 0} \,. \label{qV5_prime}
\ee

Next, we further expand \1eq{V10_step_0} around $q = 0$. The zeroth order term is easily seen to vanish, while the first nonvanishing term is given by
\be 
\qV5^\prime(q) = q^2 f_1(0)  \,, \label{V10_step_1}
\ee
where
\be 
f_1(0) := - \frac{F(q)m^2}{2}\left\lbrace A_3(0,0,0) + \left[ \frac{d}{d q^2}\left( A_3^\p(0,q,-q) - A_3^\p(q,-q,0) \right)\right]_{q = 0} \right\rbrace \,. \label{f10}
\ee

Then we relate $\Rc{10}^\prime$ to $\qV5^\prime$ by expanding \1eq{Vs_2basis} around $p = 0$. In doing so, we make extensive use of the Bose symmetry relations of \2eqs{VBose}{VBose_2}, and invoke \1eq{eq:taylor_C}. After some algebra, this procedure yields
\be 
\Rc{10}^\prime(q) = \qV5^\prime(q) + q^2 f_2(q) \,, \label{V10_step_2}
\ee
where
\be 
f_2(q) := \Cfat(q) - q^2 \left[ \frac{\partial \Rc2(p,q,r)}{\partial r^2}\right]_{p = 0} + \left[ \frac{ \partial }{\partial r^2}\left( \Rc7(r,p,q) - \Rc7(q,p,r) \right) \right]_{p = 0} \,.
\ee

Finally, combining \2eqs{V10_step_1}{V10_step_2} we obtain the announced result, \1eq{V10_goal}, by identifying $f(0) := f_1(0) + f_2(0)$.

\subsection{SDE realization}\label{subsec:V10_dyn}

Now, we show how \1eq{V10_goal} 
follows from the SDE of the three-gluon vertex. Note that, by virtue of \1eq{V10_step_2}, which is a consequence of Bose symmetry, it suffices to demonstrate \1eq{V10_step_1}.

To this end, we employ a procedure similar to that used in Section~\ref{V9SDE} to obtain $\Rc9(q)$. Specifically,  
({\it i}) we contract the vertex SDE of \fig{fig:V9diags} by $q_\alpha$; ({\it ii}) then, we isolate the tensor structure $r_\mu p_\nu$ from the result, which yields a contribution to $\qV5/(r^2p^2) + \qG{5}$; ({\it iii}) next, we multiply the result by $r^2 p^2$ and expand to lowest order in $p = 0$, thus obtaining $\qV5^\prime$; ({\it iv}) lastly, we expand $\qV5^\prime$ to lowest order around $q = 0$.

In carrying out step ({\it i}) above, we note that diagrams $(e_1)$, $(e_3)$ and $(e_4)$ of \fig{fig:V9diags} do not contribute to $\Rc{10}$, for the exact same reasons
that they do not contribute to $\Rc9$, as 
discussed in Section~\ref{V9SDE}.
Hence, we focus on diagram $(e_2)$. Moreover, after triggering the STI of \1eq{4gSTI} for the four-gluon vertex in $(e_2)$, we see that only the term  highlighted with an underbrace can contribute to $\qV5$. Hence, we are led back to \1eq{qd2}. 

Then, we carry out  step ({\it ii}). Evidently, only the term $r_\mu \mathbb{B}_2(p,q,r)$ of \1eq{pH} contributes to the tensor structure $r_\mu p_\nu$. Hence, we can write
\begin{align} 
q^\alpha (e_2)_{\alpha\mu\nu} =&\, - \frac{\lambda Z_3 F(q) \mathbb{B}_2(p,q,r) r_\mu}{p^2}\int_k\Delta^{\rho\delta}(k-p)\Delta^{\sigma\tau}(k)\g^{(0)}_{\nu\sigma\rho}(-p,k,p-k)  \nonumber\\
\times&\, \left[ V_1(-p,k,p-k) g_{\delta\tau} + V_2(-p,k,p-k)p_{\delta}p_{\tau}\right] + \ldots \,, \label{qd2_H_V10}
\end{align}
with the ellipsis denoting terms that cannot contribute to $\qV5$ because they do not contain either a $1/p^2$ or a $r_\mu$.

Then, for small $p$, we can expand \1eq{qd2_H_V10} around $p = 0$, using \1eq{eq:taylor_C}. Note that to first order in $p$, \1eq{B1_B2} implies $\mathbb{B}_2(p,q,r) = - (p\cdot q)\mathbb{A}_3(0,q,-q)$. Hence, we obtain
\begin{align} 
q^\alpha (e_2)_{\alpha\mu\nu} =&\, - F(q) (q\cdot p ) q^2 \mathbb{A}_3( 0,q,-q) \left( \frac{ r_\mu p_\nu }{r^2 p^2 } \right) \left[ 3\lambda Z_3  \int_k k^2 \Delta^2(k) \Cfat(k) \right] + \ldots \,, \label{qd2_H_V10_step1}
\end{align}
with ellipsis now including terms that are dropped in the expansion around $p = 0$.

To complete step ({\it ii}), we note that the form factor of the tensor $r_\mu p_\nu$ is $\qV5/p^2r^2 + \qG5$. Hence, invoking \1eq{mass_origin},
\begin{align} 
\frac{\qV5^{(e_2)}}{p^2r^2} + \qG5^{(e_2)} = \frac{\qV5}{p^2r^2} + \qG5^{(e_2)} = - \frac{(q\cdot p)}{r^2p^2}F(q) q^2 \mathbb{A}_3( 0,q,-q) m_{(a_1)}^2 + \ldots \,. \label{qd2_H_V10_step2}
\end{align}
Note that in the first equality, we used the fact that only $(e_2)$ contributes to $\qV5$, \ie $\qV5 = \qV5^{(e_2)}$, whereas $\qG5$ may receive contributions from other diagrams.

Proceeding to step ({\it iii}), we multiply \1eq{qd2_H_V10_step2} by $r^2 p^2$ and expand the result to the first order in $p$. Using \1eq{qV5_prime}, we find
\be 
\qV5^\prime(q) = - \frac{F(q) m_{(a_1)}^2}{2} \left[ q^2 A_3( 0,q,-q) + A_3^\p(0,q,-q) \right] \,. \label{qd2_H_V10_step3}
\ee
Note that this result is nearly identical to \1eq{V10_step_0}, differing from it only by the substitutions $m\to m_{(a_1)}$ and $A_3^\p(q)\to 0$.

Finally, we perform  ({\it iv}), \ie expand \1eq{qd2_H_V10_step3} around $q = 0$. Using \1eq{A3p_0}, we obtain
\be 
\qV5^\prime(q) = q^2 f_3(0) \,, \qquad f_3(0) := - \frac{F(0) m_{(a_1)}^2}{2}  \left\lbrace A_3( 0,0,0) + \left[\frac{d A_3^\p(0,q,-q)}{d q^2}\right]_{q = 0} \right\rbrace \,, \label{qd2_H_V10_step4}
\ee
which is \1eq{V10_step_1}, with 
$f(0) := f_3(0)$. 

As in the previous section, 
the STI results 
given by \2eqs{V10_step_1}{f10}, 
and the SDE result 
in \1eq{qd2_H_V10_step4} 
are strikingly similar; again,  
the observed discrepancy
is due to the SDE 
truncation, 
or the approximate nature of 
the STI in \1eq{4gSTI}.

\section{Conclusions}\label{conc}

The intense scrutiny of the 
correlation functions of QCD by means of 
continuous methods~\cite{Aguilar:2008xm,Fischer:2008uz,Binosi:2012sj, Aguilar:2016vin,RodriguezQuintero:2010wy,Aguilar:2002tc,Aguilar:2004sw,Horak:2022aqx}, and lattice simulations~\cite{Cucchieri:2007md,Cucchieri:2007rg,Bogolubsky:2007ud,Bogolubsky:2009dc,Oliveira:2009eh,Oliveira:2010xc,Cucchieri:2009zt},   
supports the notion that 
the gluons acquire a nonperturbative mass~\cite{Aguilar:2021uwa,Aguilar:2022thg} through the action of the celebrated 
Schwinger mechanism. 
In a non-Abelian context, 
the main dynamical characteristic of this mechanism 
is the formation of composite massless poles in the vertices of the 
theory~\mbox{\cite{Jackiw:1973tr,Jackiw:1973ha,Eichten:1974et,Poggio:1974qs,Smit:1974je,Cornwall:1973ts,Cornwall:1979hz}}. These poles display the crucial 
feature of being completely longitudinally 
coupled, a fact that guarantees the 
absence of divergences in 
(Landau gauge) 
lattice form factors.

In this article 
we have analyzed the pole content of 
the three-gluon vertex, whose role is known to be instrumental 
in the realization of the Schwinger mechanism,
accounting for the bulk of the gluon mass~\cite{Aguilar:2021uwa,Aguilar:2022thg,Papavassiliou:2022wrb,Ferreira:2023fva}. 
It turns out that the resulting structures are quite rich, 
being imposed by 
Bose-symmetry and the STI
satisfied by the 
three-gluon vertex. 
In particular, 
we have focused on the appearance and role 
of the mixed double and triple poles, 
of the type $1/q^2p^2$ and $1/q^2 r^2 p^2$, 
respectively, 
which are inert as far 
as the direct act of mass generation is concerned. 

It turns out that the mixed double poles 
are an indispensable requirement for 
the flawless completion of the STI satisfied by this vertex in the presence of an infrared finite (massive) gluon propagator. 
In fact, the STI imposes powerful constraints relating the so-called ``residue function'' to all other components entering in the STI. 
We emphasize that, at this level, 
the presence of these poles 
is dictated solely by the 
STI, and 
is not related to any particular 
dynamical realization. In that sense, it  
appears to be of general validity,
hinging only on the longitudinal nature of the poles. The picture  emerging from the
bound-state 
realization of the Schwinger mechanism, 
as captured by the vertex SDE~\mbox{\cite{Aguilar:2011xe,Ibanez:2012zk,Aguilar:2017dco,Binosi:2017rwj}}, 
satisfies the general constraints 
imposed by the STI, 
thus passing a highly nontrivial 
self-consistency check. 

As for the mixed triple pole, our analysis 
reveals that their strength is 
substantially reduced (\ie weaker than a double mixed pole), 
again by virtue of 
inescapable 
requirements imposed by the STI. 
Interestingly enough, 
the salient qualitative features of this result are recovered by the vertex SDE, 
exposing once again the 
complementarity 
between symmetry and dynamics. 

Our analysis strongly indicates 
that higher $n$-point functions functions 
(\ie Green's functions with $n$ 
incoming gluons, and $n>3$)  
will also possess an extended 
structure of poles. 
This is already seen at the level of the four-gluon vertex
$\fatg_{\alpha\mu\delta\tau}^{abts}$
($n=4$), which enters  
in the demonstration of   
Sec.~\ref{V9SDE}. In particular,  
$\fatg_{\alpha\mu\delta\tau}^{abts}$
is forced by the STI of 
\1eq{4gSTI}, namely by the 
$V$-parts of the three-gluon vertices 
appearing on the r.h.s., 
to have poles in all channels 
carrying the momenta of the 
external legs, together 
with the channels obtained by forming sums of momenta, as happens in the case
$p=q+r$. A preliminary study 
reveals that the diagrammatic 
interpretation of all these poles 
is fully consistent with the 
notions and elements introduced  
in Sec.~\ref{sec:background}. 
We hope to report the 
results of a detailed inquiry 
in the near future.

It would be clearly important to 
unravel an organizing principle 
that accounts for the 
pole proliferation 
in the fundamental vertices of QCD. 
A possible approach is 
the construction of 
low-energy effective 
descriptions of Yang-Mills theories 
with a gluon mass, in the spirit 
of the gauged non-linear sigma model
proposed by Cornwall,
see~\cite{Cornwall:1974hz,Cornwall:1979hz,Cornwall:1985bg,Cornwall:2010upa}. 
In this model, the addition of a gluon 
mass at the level of the effective 
Lagrangian is compensated by 
the presence of angle-valued 
scalar fields, which act as would-be 
Nambu-Goldstone particles. 
When the equation of motion of these scalars is solved as a power series in the gauge field, and the solution is substituted back into the 
Lagrangian, the various vertices 
acquire 
longitudinally coupled massless 
poles, see, \eg \1eq{V_Cornwall}.
A systematic comparison between the  
pole patterns  obtained within 
this model (or variants thereof) 
and those 
induced by  
the Schwinger mechanism at the 
level of the 
fundamental theory,  
as detailed here,
might afford clues on the 
structure of possible low-energy effective descriptions of QCD. 

\section{Acknowledgments}
\label{sec:acknowledgments}

The work of  A.~C.~A., B.~M.~O., and L.~R.~S. is supported by the CNPq grants \mbox{307854/2019-1}, \mbox{141409/2021-5}, and \mbox{162264/2022-4}, respectively. A.~C.~A also acknowledges financial support from project 464898/2014-5 (INCT-FNA). M.~N.~F. and J.~P. are supported by the Spanish MICINN grant PID2020-113334GB-I00. M.~N.~F. acknowledges financial support from Generalitat Valenciana through contract \mbox{CIAPOS/2021/74}. J.~P. also acknowledges funding from the regional Prometeo/2019/087 from the Generalitat Valenciana.

%

\end{document}